\documentclass[a4paper,11pt]{article}

\usepackage{jcappub} 
\usepackage{lineno}
\usepackage{longtable}
\usepackage{supertabular}
\usepackage{booktabs}
\usepackage{multirow}
\usepackage[utf8]{inputenc}   
\DeclareUnicodeCharacter{2212}{-}
\usepackage{hyperref}
\usepackage[T1]{fontenc}
\usepackage{appendix}
\usepackage{amsmath,amssymb}
\usepackage[makeroom]{cancel}
\usepackage{graphicx}
\usepackage{dcolumn}
\usepackage{booktabs}
\usepackage{multirow}
\usepackage{bm}
\usepackage{hyperref}
\usepackage{braket}
\usepackage{soul}
\usepackage{subcaption}

\usepackage{xcolor}     

\title{\boldmath $Q$-balls, neural networks and galaxy rotation curves}

\author[a,1]{Alexandre M. Pombo\note{Corresponding author.},}
\author[b]{Lorenzo Pizzuti}
\author[b]{and Alessandra di Giacomo}

\affiliation[a]{CEICO - Institute of Physics of the Czech Academy of Sciences}
\affiliation[b]{Dipartimento di Fisica “G. Occhialini”, Universit\`a degli Studi di Milano Bicocca, Piazza della Scienza 3, 20126 Milano, Italy}

\emailAdd{pombo@fzu.cz}
\emailAdd{lorenzo.pizzuti@unimib.it}
\emailAdd{a.digiacomo12@campus.unimib.it}

\abstract{Can a dynamically robust (\textit{aka} stable) $Q$-ball reproduce the rotation curve of a disk galaxy? In an astrophysical environment, $Q$-balls are non-topological solitons that are transparent and only perceived by their gravitational effects. Traditionally, scalar $Q$-balls are modelled with a polynomial potential, but axion-like periodic potentials are also expected to support such solitonic configurations. In the presence of angular momentum, $Q$-balls acquire a toroidal structure with a central density void, qualitatively resembling the axially-symmetric structure of disk galaxies. Motivated by this similarity, we investigate whether rotating scalar $Q$-balls can reproduce the observed rotation curves of disk galaxies. In this work, we use a recently developed hybrid numerical framework that combines a high-accuracy pseudo-spectral method with a physics-informed neural network approach to construct both static and rotating $Q$-ball solutions. We assess their ability to act as the dark matter halos in galaxies by fitting the observed rotation curves of a sample of disk galaxies from the SPARC catalogue. Our simplified model provides an overall good agreement with observational data, and a reasonable fit when compared to standard dark matter profiles such as the Navarro-Frenk-White; we have further found an average constraint on the scalar field particle's mass $m\sim 10^{-27}$ eV, in agreement with similar galactic-scale soliton solutions.}

\begin{document}

\maketitle

\flushbottom

%
\section{Introduction}\label{S1}
%
    The study of galaxy rotation curves (RCs) has long provided compelling evidence for the existence of extended non-luminous mass components in spiral and disk galaxies (\textit{e.g.} \cite{Rubin1978,Bosma1981}). Such measurements show that the orbital velocities of gas and stars remain roughly constant at large radii, rather than falling off as expected from the luminous matter distribution alone (\textit{e.g.}~\cite{lelli_2016_sparc}) -- this is known as the rotation curve problem. Such goes against the standard Newtonian laws and the expected Keplerian decline. Assuming that gravity is described by standard general relativity (GR), the flattening of rotation curves can be explained through the inclusion of an additional matter contribution that must account for $\sim 80\%$ of the total mass contribution. This became known as dark matter (DM), with galaxies' rotation curves being some of the first and most reliable arguments in favour of a non-baryonic DM. 
    
    Due to the complex physics involved in galaxy formation and evolution, RCs exhibit a quite wide variety of behaviours (\textit{e.g.}~\cite{Oman2015})\footnote{Dwarf-galaxy RCs often point towards more cored inner DM profiles, challenging the cuspy halos of cold DM-only simulations \cite{Oh2015}}. Recent works highlight that non-equilibrium gas motions and systematics can bias RC inferences, partly explaining the observed diversity \cite{Sands2024}, while high-redshift data show steeper outer declines than in local spirals \cite{NelsonWilliams2024}. Cosmological hydrodynamics reproduce systematic RC trends with mass and morphology \cite{Jeong2025}, but open questions remain. Proposed solutions span \textit{i)} baryonic feedback that reshapes inner halos, \textit{ii)} particle physics alternatives such as self-interacting DM \cite{TulinYu2018}, and \textit{iii)} modified gravity frameworks (\textit{e.g.}, MOND~\cite{Milgrom1983}; Refracted Gravity \cite{Matsakos16}); those, however, still struggle to fit observations at larger scales (\textit{e.g.} \cite{Pizzuti25,Famaey25}) -- for a review of small-scale challenges, see also \cite{Bullock_2017}.
    
    Recently, models considering ultra-light bosonic DM (``fuzzy'' DM) have been gaining traction in both astrophysical and cosmological scenarios \cite{Marsh_2022,dellaMonica23,Salasnich_2025}. In this context, DM is described by a fundamental field with a distinctive tiny mass $m\sim 10 ^{-22}-10^{-27}$ eV, much lighter than standard cold-DM candidates (few GeV–100 TeV). Such a tiny mass corresponds to a de Broglie wavelength on galactic scales, producing a quantum-pressure term that counteracts gravitational self-attraction and leads to a self-gravitating Bose-Einstein condensate-like soliton known as a boson star \cite{jetzer1992boson,kleihaus2005rotating,herdeiro2017asymptotically,pombo2023coupled,brito2016proca} (see \cite{shnir2023boson} for a review).
    
    When self-interactions of the field dominate over gravity, non-topological solitons known as $Q$-balls arise~\cite{coleman1985q,volkov2002spinning,lee1989gauged,loiko2018q,axenides2000dynamics} -- localised, self-reinforcing configurations stabilised by a conserved Noether charge and held together by attractive self-interactions rather than gravity. These $Q$-ball solutions, which share many properties with boson stars in the Newtonian limit, are the focus of the present project.
    
    In a recent series of works~\cite{Mourelle24,Mourelle25} the authors investigated the possibility of boson stars to describe the observed rotation curve of spiral galaxies. In their study, they observed that a combination of scalar and vector field  boson stars is able to reproduce the observed rotation curve of several spiral galaxies.

    In this paper, we study whether the RC of disk galaxies can be described by a single, rotating, scalar field $Q$-ball that natively reproduces the axial symmetry of the galactic systems. In this case, the fact that solutions are governed by self-interactions -- rather than self-gravitation -- allows a wider range of stable rotating solitonic solutions that do not exist in the single non-self-interacting spinning scalar boson star case.
    
    The standard construction of scalar $Q$-balls relies on a sextic self-interaction polynomial potential. Recently, an axion-like potential has been gathering interest in the construction of scalar boson stars~\cite{guerra2019axion,delgado2020rotating,pombo2023sun} whose properties mimic the standard model at high frequencies while significantly deviating as one progresses in the domain of existence. One of the emerging properties is the existence of an additional stable branch \cite{herdeiro2021imitation} which was not previously present and gives the possibility of further astrophysically relevant boson star configurations to exist. Such an axion-like potential is also expected to yield $Q$-ball solutions. The self-interaction potential expands as a sextic self-interaction polynomial potential for small field amplitudes, while once again deviating as the field increases. With this behaviour, we expect axion-like $Q$-balls to possess different properties than their simple polynomial counterpart. In this work, we will build such axion-like $Q$-ball configurations, analyse their properties, and compare the RC signature they yield against the standard polynomial $Q$-ball.
    
    For both potential types, we will fit the obtained $Q$-ball RCs to data from a set of high-precision RCs of nearby galaxies within the Spitzer Photometry \& Accurate Rotation Curves (SPARC) catalogue \cite{lelli_2016_sparc}, and we infer constraints on the scalar field' particle mass. We further compare the fitted curves to those obtained by assuming the well-established Navarro-Frenk-White (NFW) model as DM halo profile. 

    To numerically construct the solutions, we introduce a newly in-house developed solver that hybridises the pseudo-spectral method (PSM) with the physics-informed neural network (PINN) framework. In this approach, the PSM basis functions are embedded as neural activation functions, preserving spectral accuracy while incorporating the flexibility of machine learning. We refer to this hybrid solver as \texttt{SpectralPINN}. Although this type of hybridisation remains unexplored, both PSM and PINNs are well-established techniques individually. Notably, both methods ultimately reduce to a weight-optimisation problem, making them naturally compatible.

    The loss function enforces the $Q$-ball's equations of motion and required boundary conditions by minimising their residual, ensuring that the obtained solutions remain physically consistent. This framework leverages the complementary strengths of both approaches, combining the spectral precision of PSM with the adaptability of PINNs, while mitigating the black-box nature of standard neural architectures. In practice, \texttt{SpectralPINN} learns the coefficients (amplitudes) of the spectral basis functions directly through a PINN, solving the governing system of PDEs under the imposed boundary conditions. The resulting model yields a compact, storage-efficient, closed-form analytical representation of the fields -- in the form of the PSM basis amplitudes -- rather than opaque network outputs.

    Unless otherwise stated, throughout the text we define the physical units as: $c = 63241$ $\rm{AU} \cdot \rm{yr}^{-1}$, $G \simeq 39.748$ $\rm{AU}^3 \cdot M_{\odot}^{-1} \cdot \rm{yr}^{-2}$, $\rm{M_{P}} \simeq 1.094 \cdot 10^{-38} \ M_\odot$. For the numerical construction of the $Q$-balls, we consider geometrized units with $4\pi G=1=4\pi\epsilon_0$. The signature of the spacetime is $(-,+,+,+)$. 

    The paper is organised as follows. Section~\ref{S2} introduces the $Q$-ball model, deriving the corresponding field equations, boundary conditions, and relevant physical quantities. Section~\ref{S3} presents the \texttt{SpectralPINN} method, which is applied in Section~\ref{S4} to obtain and analyse the $Q$-ball solutions and their domain of existence. Section~\ref{S6} describes the conversion from geometrised to physical units and the computation of the RCs, while Section~\ref{S5} introduces the SPARC dataset. Comparison between the galaxy dataset and expected RC is done in Section~\ref{S7}. Finally, Section~\ref{S8} summarises the main conclusions.
%
\section{$Q$-balls model}\label{S2}
%
	In its simplest form, a $Q$-ball is constructed in a field theory of a complex scalar field, $\Phi$, in which the Lagrangian is invariant under a global \textbf{U(1)} symmetry and is given by
	\begin{equation}\label{E2.1}
	   \mathcal{L}=\partial _\mu \Phi \partial ^\mu \Phi ^* -U(|\Phi |)\ .
	\end{equation}
	The global symmetry of the Lagrangian under an \textbf{U(1)} transformation of the complex field, $\Phi \rightarrow \Phi e^ {a i}$,  gives rise to a conserved Noether current
    \begin{equation}
     j_\mu = i \left( \Phi^* \partial_\mu \Phi - \Phi \, \partial_\mu \Phi^* \right)\ ,
    \end{equation}
    and the associated conserved charge,
    \begin{equation}
     Q = \int d^3x \, j^0 = i \int d^3x \left( \Phi^* \dot{\Phi} - \Phi \, \dot{\Phi}^* \right)\ ,
    \end{equation}
    which corresponds to the conserved particle number.	The fundamental $Q$-ball solutions are minima of the energy for a given $Q$. Since $\Phi$ should depend on time for $Q$ to be non-vanishing~\cite{derrick1964comments}, following the literature, we assume an ansatz that contains both a harmonic time and azimuthal dependence
	\begin{equation}\label{E2.4}
	   \Phi(t,r,\theta,\varphi)\equiv \phi (r,\theta) e^{i(m\varphi-\omega t) }\ , 
	\end{equation}
	where $\phi (r,\theta )$ is a real function that describes the scalar field's spatial amplitude; $\omega$ is the scalar field frequency -- which measures the fundamental energy state of the bosonic particles --, and $m$ is the azimuthal harmonic index. 

	Since gravity is negligible compared to the field's self-interaction,  we consider an asymptotically flat Minkowski spacetime described by the line element,
	\begin{equation}
	   ds^2 = -dt^2 +dr^2 +r^2 (d\theta ^2 +\sin ^2 \theta d\varphi ^2 )\ .
	\end{equation}		
	Due to the choice of the harmonic time dependent ansatz \eqref{E2.4}, neither the self-interaction potential, $U(|\Phi |)\equiv U(\phi )$, nor the energy-momentum tensor
	\begin{equation}
	   T_{\mu \nu} =\partial _\mu \Phi \partial _\nu \Phi ^* +\partial _\nu \Phi \partial _\mu \Phi ^* -g_{\mu \nu} \mathcal{L}\ ,
	\end{equation}
	depends explicitly on time. The energy distribution is therefore stationary, and the total energy is
	\begin{equation}\label{E2.7}
	   E=\int d^3 x\, T_{tt} = 2\pi \int _0 ^{+\infty } dr r^2 \int _0 ^\pi d\theta\, \sin \theta\,  \left[ \omega ^2 \phi ^2 +(\partial_r \phi)^2+\frac{(\partial_\theta \phi)^2}{r^2}+\frac{m^2 \phi ^2}{r^2 \sin ^2 \theta}+U(\phi) \right]\ , 
	\end{equation}
    while the Noether charge is,
	\begin{equation}\label{E2.8}
	   Q= 2\omega \int \phi ^2 d^3 x = 4\pi \omega \int _0 ^{\infty} dr\, r^2 \int _0 ^\pi d\theta\, \sin \theta\, \phi ^2 \ .
	\end{equation}
	Since we are considering rotating solutions, another quantity of interest is the $Q$-ball's angular momentum, $J$, which can be shown to be quantised and related to the Noether charge as
	\begin{equation} \label{eq:angular_momentum}
     J=\int d^3 x \,  T_{t\phi }  = m Q\ .
	\end{equation}
%
    \subsection{Axion-like potential}\label{S2.1}
%
    For the existence of a self-sustaining soliton (following Coleman~\cite{coleman1985q}), the model must possess a well-defined ground state, given by the existence of a global minimum of the non-negative self-interacting potential $U$ at $\Phi =0$, where $U(0)=0$ and $U'(0)=0$, and a positive mass for small fluctuations $\mu^2 \equiv U''(0) >0$. Additionally, there must exist a range of $\omega$ values for which solutions exist, with the maximum frequency given by
    
        \begin{equation} \label{eq:condition1}
		   \omega ^2 <\omega _+ ^2 \equiv \frac{1}{2}\frac{d^2 U}{d\phi ^2}\bigg| _{\phi = 0}\ ,
        \end{equation}
    while the requirement of a region with an attractive self-interaction potential yields the minimum value of the frequency
        \begin{equation} \label{eq:condition2}
		   \omega ^2 > \omega _- ^2 \equiv \text{min} _\phi \frac{2 U}{\phi ^2}\ ,
        \end{equation}
    with $\omega _+ > \omega _-$. Finally, in order to possess a localised/finite energy soliton, $U$ must grow faster than $\phi^2$ for large $\phi$, ensuring the energy is bounded from below. 

    As already mentioned in Section~\ref{S1}, the standard $Q$-ball potential is a simple sextic-order potential in $\phi$, namely 
        \begin{equation}\label{E2.12}
	       U_P(\phi ) =\mu^2 \phi ^2 +\beta\phi ^4 +\lambda \phi ^6 \ ,
	    \end{equation}
	which, although not renormalizable, satisfies the necessary conditions. A standard choice for the potential parameters is $\mu = 1.0\,,\ \beta =-1.8\, , \lambda=1.0$. Let us now focus on the massive axion-like potential first introduced in \cite{guerra2019axion}
	    \begin{equation}\label{E2.13}
	       U_A (\phi) = \frac{2 \mu ^2 \alpha ^2 }{B}\left[ 1-\sqrt{1-4B \sin ^2 \left(\frac{\phi}{2\alpha}\right) }\ \right]\ ,
	    \end{equation}
	where $B=\frac{z}{(1+z)^2 }\approx 0.22$ and $z= m_u /m_d \approx 0.48 $ is the mass ratio of the up and down quarks, while the second term in the $U_A$ potential is the standard QCD Axion potential, to which was added the first constant term to ensure $U_A (0)=0$ and hence asymptotically flat solutions. The axion-like potential, just like its polynomial counterpart, is non-renormalizable\footnote{In \(3+1\) dimensions, power-counting renormalizability requires that the Lagrangian contain only a finite set of operators with canonical mass dimension \(\leq 4\) (\textit{e.g.}\ \(\phi^2,\ \phi^4\) for a real scalar). Expanding \(U_A\) around \(\phi=0\) generates an infinite tower of higher-dimension interactions \eqref{E2.14}, so that, in the Dyson sense, an infinite number of counter-terms would be required and hence the theory is non-renormalizable.}.

	Expanding the $U_A$ potential around the minimum $\phi = 0$, we obtain
	    \begin{equation}\label{E2.14}
	       U_A (\phi) \approx \mu ^2\phi ^2+\frac{(3 B-1)}{12 \alpha ^2}\mu ^2 \phi^4 +\frac{15 B (3 B-1)+1}{360 \alpha ^4} \mu ^2 \phi^6\,
	    \end{equation}
     which possesses the same functional form as the standard $U_P$. For the sake of comparison between the two potentials, one must impose $\mu = 1$ for both cases. This sets the particle's mass with which the $Q$-ball's quantities are scaled with respect to. Concerning the two other possible parameters, as already mentioned, $B$ is fixed by the ratio between the up and down quark, $B=0.22$, which is compatible with a bounded $Q$-ball solution ($0<B<0.25$). This leaves a single free parameter that will allow the potential to obey the required conditions and reproduce the quartic term of $U_P$. From the second quadratic term in the expansion, 
        \begin{equation} \label{eq:parameters}
         \frac{(3 B-1) \mu ^2 \phi^4}{12 \alpha ^2}= \beta\ , \qquad \Rightarrow \qquad  \alpha = \pm 0.1255\ .
        \end{equation}
    Since the potential is only sensitive to $\phi/\alpha$, we choose $\alpha>0$, which will also make $\phi >0$. Observe that imposing $\alpha=0.1255$ gives a sextic order expansion term $\frac{15 B (3 B-1)+1}{360 \alpha ^4} \mu ^2=-0.3422$, which is negative and does not obey Coleman's conditions; the potential is not stabilising. However, this is merely an expansion of a sinusoidal potential and does not provide the final conclusion. The full $U_A$ potential is safe and provides a non-empty $\omega$ existence window. 
	
	From the Coleman's potential conditions, the existence of a maximum frequency provided by \eqref{eq:condition1} gives $\omega _+ = \mu = 1$ for both potentials\footnote{As it will be discussed in Section~\ref{S4}, as the solution gets closer to the maximum frequency $\omega \to \omega_+$ the maximum value of the scalar field amplitude decreases, $\phi \to 0$, and the leading contribution to the potential comes from the mass term, which is common for both potentials.}. For the minimum scalar field frequency, \eqref{eq:condition2} gives $\omega_- = 0.6164 $ for $U_P$ and $\omega _- = 0$ for $U_A$, both obeying the non-empty $\omega$-window condition. Observe that, in the spherically symmetric case, $m=0$, while the polynomial potential gives solutions that are always complex $\omega>0$, the axion-like potential seems to allow the existence of real solutions by having $\omega = 0$, however, this is a limiting degenerate solution that our soliton will not be able to reach since the $Q$-ball requires a non-zero Noether charge to exist and hence, by \eqref{E2.8} a non-zero $\omega$.
    
%
	\subsection{Equations of motion}	
%
	The model's Lagrangian \eqref{E2.1} with the scalar field ansatz \eqref{E2.4} yields a single second-order PDE
	   \begin{equation} \label{E2.16}
	    r^2 \partial_r ^2 \phi +2 r \, \partial_r \phi+ \partial_\theta ^2 \phi +\cot \theta\, \partial_\theta \phi-\frac{m^2}{\sin ^2 \theta}\phi+r^2\bigg(\omega^2 \phi+\frac{dU}{d\phi^2}\phi \bigg)= 0\ ,
	   \end{equation}
	Regularity of the spinning solution at the poles requires $\phi (r,\theta =0)=\phi (r,\theta =\pi)=0$. Near the origin, the field behaves as:
	   \begin{equation}\label{E2.17}
	    \phi (r,\theta) \approx \phi_1 \, r^m \sin ^m\theta+\mathcal{O}(r^2)\ ,
	   \end{equation}
	while at infinity, $r\rightarrow + \infty$, energy finiteness of the solutions requires $\phi (r\to \infty,\theta)=0$, and the field can be approximated by
	   \begin{equation}
	    \phi (r) \approx Q_m(\theta) \frac{e^{-\sqrt{\mu ^2-\omega ^2}\, r}}{r}\ ,
	   \end{equation}
    with $Q_m (\theta)\propto Q_s \sin^m \theta$, and $Q_s$ and $\phi _1$ two integration constants. Observe that, while our numerical procedure does not demand the field's asymptotic expansions, these help us build and analyse the behaviour of the $Q$-ball solutions.
    
    At last, in this work, we shall solely focus on solutions that are symmetric with respect to a reflection along the equatorial plane, $\theta =\pi/2$. These are even parity solutions. Typically, this is the case for fundamental solutions, but there can be odd-parity excited states. This choice means that one only has to consider the range $0\leqslant \theta \leqslant \pi /2$.

	To test our numerics, besides the numerical stability tests included during the computation, one can also use the virial relation obtained through Derrick's scaling argument~\cite{derrick1964comments}. Following the strategy developed in \cite{pombo2025virial,oliveira2023convenient,herdeiro2022deconstructing,herdeiro2021virial}, the virial identity of a spinning $Q$-balls solution under a non-negative self-interaction potential $U(|\phi|)$ is given by

	   \begin{equation}\label{E2.19}
	    \int _0 ^{+\infty}  dr \ r^2 \int _0 ^\frac{\pi}{2}  d\theta \sin\theta \bigg[ r^2 \sin \theta (\partial _r \phi)^2 + \sin \theta (\partial _\theta \phi ) ^2 +\frac{m^2}{\sin \theta} \phi ^2+3r^2\sin \theta \big( U-\omega ^2 \phi ^2\big)\bigg] =0 \ ,
	   \end{equation}
	where the dispersiveness of the scalar field, $\omega$, is necessary to counter-balance the non-negative self-interaction potential, $U$.

    In what follows, for the computation of the numerical solutions we will assume a unit mass term for the field, $\mu = 1$; this is a standard procedure (\textit{e.g.} \cite{Mourelle24}), equivalent of rescaling the equation of motion in units of $\mu$, which will be referred as “code” units. We will then reintegrate the full physical units in Section \ref{S6} for the computation of the RC.
	
%
\section{SpectralPINN}\label{S3}
%
    To numerically solve the single PDE \eqref{E2.16} that describes the scalar $Q$-ball, we will resort to an in-house developed solver, \texttt{SpectralPINN}, that merges the PSM (based on \cite{blazquez2024quasinormal}) with the PINN strategy (based on \cite{luna2023solving}). To perform the computation, the method expands the unknown function $\phi (r,\theta)$ into the pseudo-spectral basis 
    \begin{equation}\label{E3.1}
     \phi (r,\theta)=\sum_{i,\,j}^{N,\, L} a_{i,j}\ T_i \big(x(r)\big)\, P_j ^m \big(y(\theta)\big)\ ,
    \end{equation}
    where $T_i$ are the Chebyshev polynomials of the first kind and $P_j^m$ are the associated Legendre functions and $N,\,L$ are the maximum basis order in our expansion of the Chebyshev and Legendre basis, respectively. Since the Chebyshev (Legendre) polynomials are only defined in the region $[-1,1]$, for unbounded radial (bounded angular) domains, we compactify it to $x \in [−1, 1]$ ($y \in [−1, 1]$) with a map $r = r(x)$ $\big(\theta = \theta (y)\big)$ given by \eqref{E3.3}. The expansion \eqref{E3.1} effectively transforms the differential equation into algebraic relations among the amplitudes $a_{ij}$. In a standard PSM, $a_{ij}$ follows from solving a (non-)linear algebraic system obtained by enforcing the PDE \eqref{E2.16} on a collocation grid; accuracy increases exponentially with the number of modes for smooth solutions.

    Alternatively, \eqref{E3.1} can be viewed as an optimisation problem: the amplitudes (weights), $a_{ij}$, are those that minimize the residual of the field equation \eqref{E2.16} while obeying boundary/regularity conditions and, when useful, the physical constraints \eqref{E2.19}. PINNs are well-suited to such residual minimisation, typically representing the solution with fully connected feed-forward neural networks.
    
\begin{figure}[h!]
	   \centering
	   \includegraphics[scale=0.9]{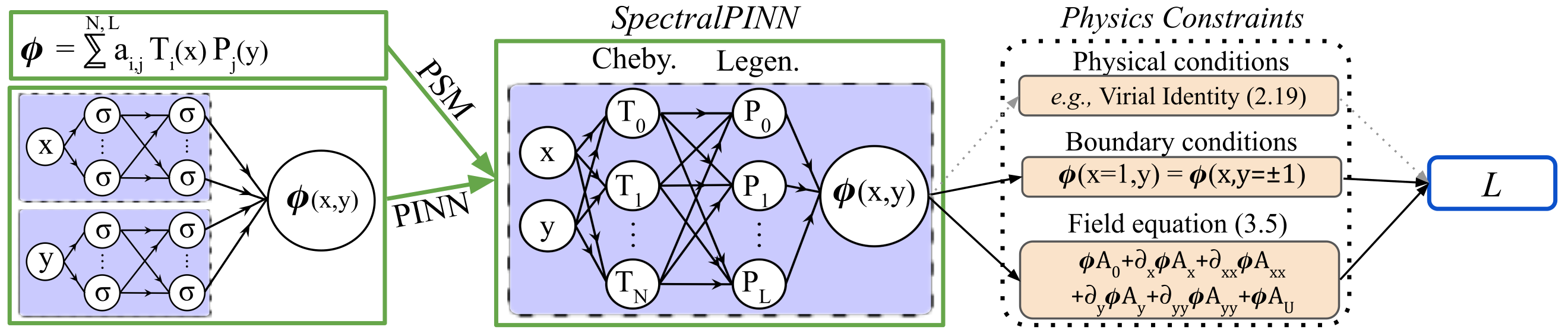}\\
	   \caption{Schematic representation of the \texttt{SpectralPINN} network architecture, where each neuron corresponds to a basis function, with each layer associated with a coordinate dependence. Scheme inspired by \cite{luna2023solving}.}
	   \label{F1}
    \end{figure}

    Our \texttt{SpectralPINN} is a hybrid: it retains the PINN training loop (stochastic/gradient-based residual minimisation) but replaces generic activations with \emph{spectral activations} that encode the problem’s symmetry. Concretely, the network’s ``layers'' correspond to the two separable coordinates and use the basis functions themselves as activations; the learnable parameters are precisely the spectral amplitudes, $a_{ij}$ (see Fig.~\ref{F1} for a schematic representation of the method). While this sacrifices PINNs’ flexibility, it gains efficiency and inductive bias for solutions compatible with the chosen symmetry.

    Following PSM practice, the basis is evaluated on grids matched to their orthogonality measures. We use a Gauss-Lobatto grid\footnote{These are the extrema of the Chebyshev polynomial \( T_N(x) \), and include the endpoints \( -1 \) and \( 1 \).} for the Chebyshev part (radial/compactified coordinate $x$) and a uniform grid for $y\equiv\cos\theta$ (Legendre part):
    \begin{align}\label{E3.2}
      x_k &=\cos\!\left(\frac{k\pi}{N_x}\right)\ ,\qquad k=0,\, \dots\, ,N_x\ ,\nonumber\\
      y_q &= -1+\frac{2q}{N_y}\ ,\qquad \ \,q=0,\, \dots\, ,N_y\ ,
    \end{align}
    with $N_x\geqslant N$ and $N_y\geqslant L$, the number of collocation points for the Chebyshev and associated Legendre polynomials, respectively. The radial and angular coordinates are compactified as:
    \begin{equation}\label{E3.3}
     x=\frac{r-1}{1+r}\ , \qquad \qquad y=\cos \theta\ .
    \end{equation}

    The resulting compactified equation on $\phi\equiv \phi \big(x(r),y(\theta)\big)$ comes as\footnote{In the compactified coordinates the boundary conditions become $\phi (x=1,y) = \phi (x, y= \pm 1)=0$.},
    \begin{align}
     4\, \phi &\left[m^2 (x-1)^2+(x+1)^2 (y^2-1)\omega^2\right] +\,(y^2-1)\bigg\{(x+1)\bigg[(x+1)\Big((x-1)^4 \partial_x ^2 \phi+2\,\frac{dU}{d\phi ^2}\Big)\nonumber \\
     &+2 (x-1)^4 \partial_x \phi\bigg] -4 (x-1)^2 (y^2-1)\, \partial_y ^2\phi -8 (x-1)^2 y \partial_y \phi\bigg\} = 0 \ .
     \label{E3.4}
    \end{align}
    where \eqref{E3.4} was rearranged to avoid any divergence term. Namely, we can rewrite the field equation as a function of the parameterising functions $A_i (x,y)$: 
    \begin{equation}
     \phi\ A_0+\partial_x\phi\ A_x+\partial_y\phi\ A_y+\partial_x ^2\phi\ A_{xx}+\partial_y ^2\phi\ A_{yy}+U\, A_U = 0\ ,
    \end{equation}
    with,
    \begin{align}
        A_U &= 2\, (x+1)^2 \left(y^2-1\right)\ ,\qquad \quad \qquad A_0 = 4 \Big[m^2 (x-1)^2+(x+1)^2 \left(y^2-1\right) \omega ^2\Big]\ ,\nonumber\\
        A_x &=2\, (x-1)^4\, (x+1) \left(y^2-1\right)\ , \quad\ \ \, A_{xx} =(x-1)^4\, (x+1)^2 \left(y^2-1\right)\ ,\nonumber\\
        A_y &=-8\, (x-1)^2\, y \left(y^2-1\right)\ , \qquad \quad \ \ \,A_{yy}=-4\, (x-1)^2\, \left(y^2-1\right)^2\ .
    \end{align}
    Since the evaluation points are defined \textit{a priori}, and both the basis functions depend solely on the coordinate points, we can pre-compute: \textit{i)} Chebyshev values, $T_i$, and differentiation matrices $T_x$, $T_{xx}$ (first/second $x$-derivatives on $x_k$); \textit{ii)} the  values of the associated Legendre functions $P^{m}_j(y_q)$ and their $y$-derivatives $P_y$, $P_{yy}$ on $y_q$; \textit{iii)} coordinate-dependent terms that emerge in the compactified field equation, $A_i$. Observe that, due to the nature of the basis functions and their recurrence relations, all the derivatives are known and computed analytically and do not require any additional differentiation approximation routine, further simplifying the computation and increasing the accuracy. These pre-computations enable us to avoid redundant computations at each training cycle, thereby significantly accelerating the training process.
    Ultimately, each training cycle reduces to the reconstruction of the function and its derivatives by the neural network, along with the respective evaluation of the weights (which we want to optimise) using the equation \eqref{E3.4}. The resulting residual is then used as the primary quantity we want to optimise, and hence serves as the main loss function component, $L_{PDE}$. Additionally, since we want to guarantee that the boundary conditions are also obeyed, these are included in the final loss function.
    \begin{equation}
        L = L_{PDE}+\beta L_{BC}\ ,
    \end{equation}
    with $\beta$ the parameter that normalises the importance of the boundary conditions' loss to the respective percentage of points associated with it, in order to avoid an over-importance of the boundary conditions with respect to the main bulk PDE evaluation. At last, it is important to point out that due to the inherited divergence at the boundaries ($y=\pm 1$), coming from the associated Legendre functions, instead of using the direct relation for the derivatives, we use the following relations:
    \begin{align}
        R_l ^m (y) &=(y^2 -1) \frac{d P_l ^m (y)}{dy} = l y P_l ^m (y)- (l+m)P_{l-1} ^m (y)\ ,\nonumber\\
        Q_l ^m (y) &= (y^2 -1)^2 \frac{d^2 P_l ^m (y)}{dy ^2} = 2 y (y^2 -1) \frac{d P_l ^m (y)}{dy}-\Big[l(l+1) (y^2 -1)-m^2\Big] P_l ^m(y) \ .  
    \end{align}
    Where the PDE's parameterizing $y$-derivative function's are adapted to absorb the $(y^2-1)$ dependence,
    \begin{equation}
        A_y = - 8 (x-1)^2 y\ , \qquad \qquad A_{yy}= -4 (x-1)^2\ .
    \end{equation}
    Finally, the associated Legendre functions and respective derivatives have a well-defined parity with respect to a reflection in the $y=0$ ($\theta = \pi/2$) plane. Since we are only interested in the even parity $Q$-ball solutions, all the odd parity associated Legendre functions do not contribute to the expansion. This implies that, in the construction of the network, only the even parity basis terms need to be considered, thereby reducing noise, improving precision, and accelerating convergence. Observe that in the absence of rotation (\textit{aka} spherical symmetry, $m=0$), only the monopolar associated Legendre function is relevant and the expansion reduces to an expansion into the Chebyshev basis.
    
    Ultimately, the problem reduces to optimising a neural network, where each neuron serves as a basis function in our expansion, with the amplitude of this basis being the weight we aim to determine, ensuring it satisfies the PDE and boundary conditions. On the neural network and training procedure technical details, the \texttt{SpectralPINN} solver uses a mean-square error loss function $MSE=n^{-1} \sum_i^n (k_i-\hat{k}_i)^2$, where $\hat{k}_i$ is the expected result, which in the \eqref{E2.16} case, $\hat{k}_i$ is always zero. To minimise the loss function, the code resorts to the Adaptive Moment Estimation with Weight Decay (\texttt{AdamW}) optimiser with a \texttt{ReduceLROnPlateau} learning rate scheduler that decreases the learning rate by $10\%$ whenever the loss function is not improved for a ``patient'' of $100$ epochs. The training process is stopped whenever the loss function reaches a value smaller than $10^{-8}$, which corresponds to a relative global error and virial identity of $10^{-11}$ and $10^{-6}$, respectively. 

    To cross-check our solutions, we also solved the spherically symmetric configuration ($m=0$) -- a single second-order ODE -- through an in-house developed, parallelised, adaptive step $5(6)^{th}$ order Runge-Kutta algorithm with a secant shooting strategy to implement the proper boundary conditions~\cite{pombo2025black}. The method yields a local error of $10^{-15}$ and a shooting tolerance of $10^{-9}$, virial identity $10^{-6}$. The solution for the $m=1$ axially symmetric configuration is obtained through a professional program package, the \texttt{CADSOL/FIDISOL}, which uses a finite difference method for integration and a Newton-Raphson strategy to implement the proper boundary conditions. The local error obtained by the method is of the order of $10^{-3}$ and the virial identity $10^{-4}$.
%
\section{$Q$-balls solutions}\label{S4}
%
    With the model and numerical procedure established, we compute the numerical solutions for both spherical and spinning $Q$-balls. Observe that, while the polynomial potential $Q$-balls are a well known and established result \cite{loiko2018q,coleman1985q,kleihaus2005rotating,volkov2002spinning,heeck2021understanding,tsumagari2008some}, solutions in the axion-like potential ($U_A$) are less explored; accordingly, we focus on $U_A$ and use $U_P$ as a baseline for comparison. 

    \begin{figure}[htbp]
        \centering
    
        \begin{subfigure}[b]{0.49\textwidth}
            \centering
            \includegraphics[width=\textwidth]{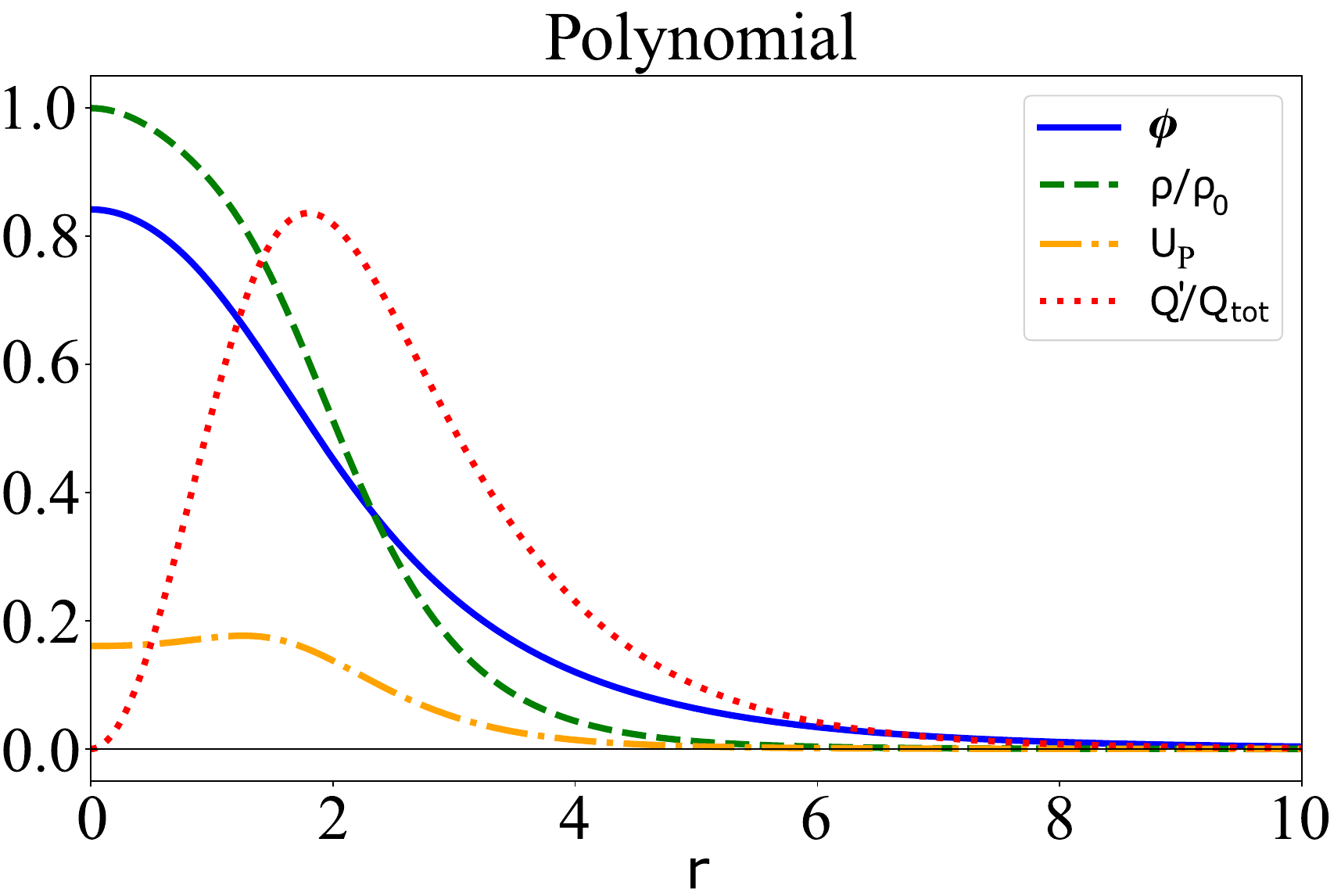}
        \end{subfigure}
        \hspace{0.0\textwidth}  
        \begin{subfigure}[b]{0.49\textwidth}
            \centering
            \includegraphics[width=\textwidth]{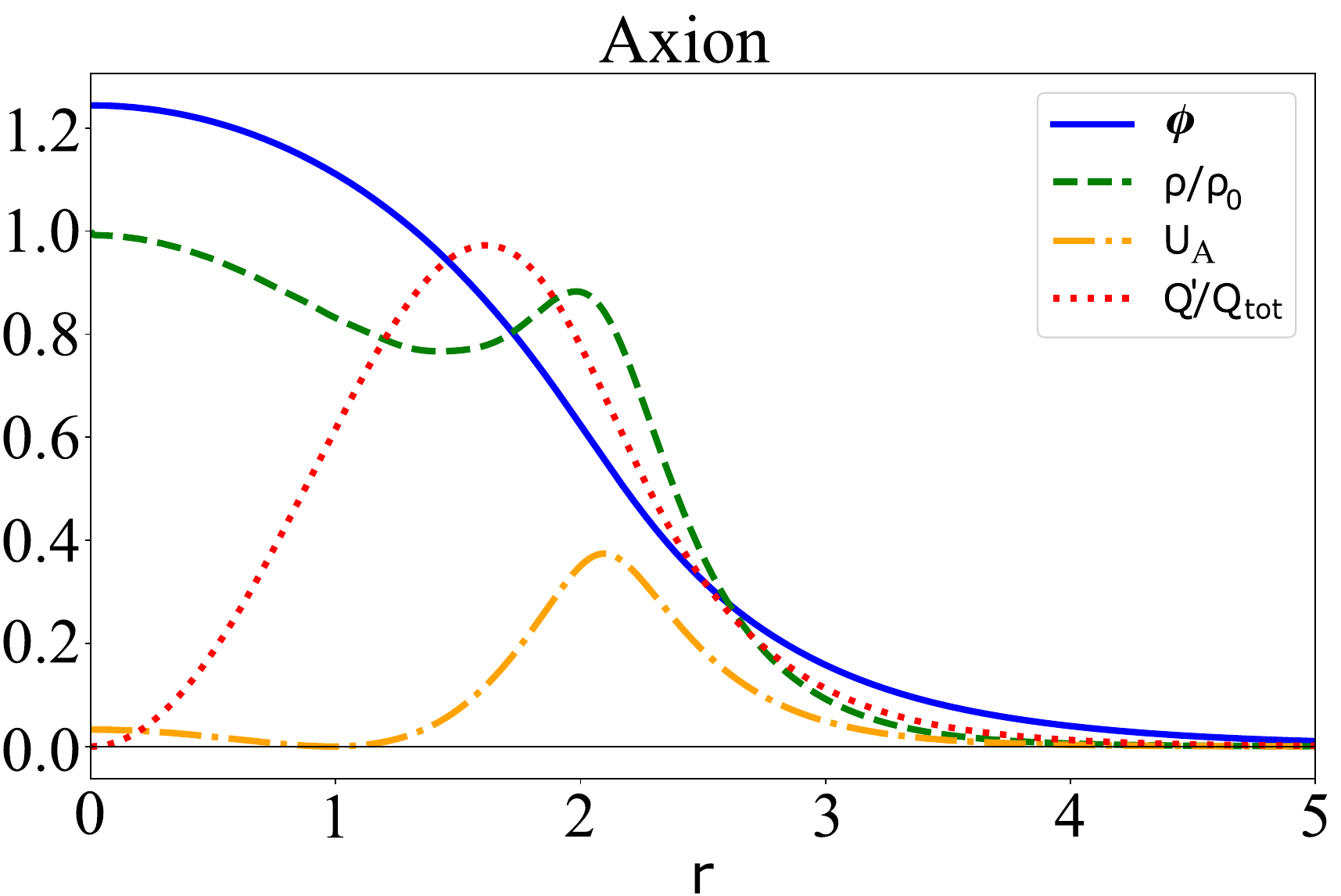}
        \end{subfigure}

        \vspace{5mm}
        \begin{subfigure}[b]{0.49\textwidth}
            \centering
            \includegraphics[width=\textwidth]{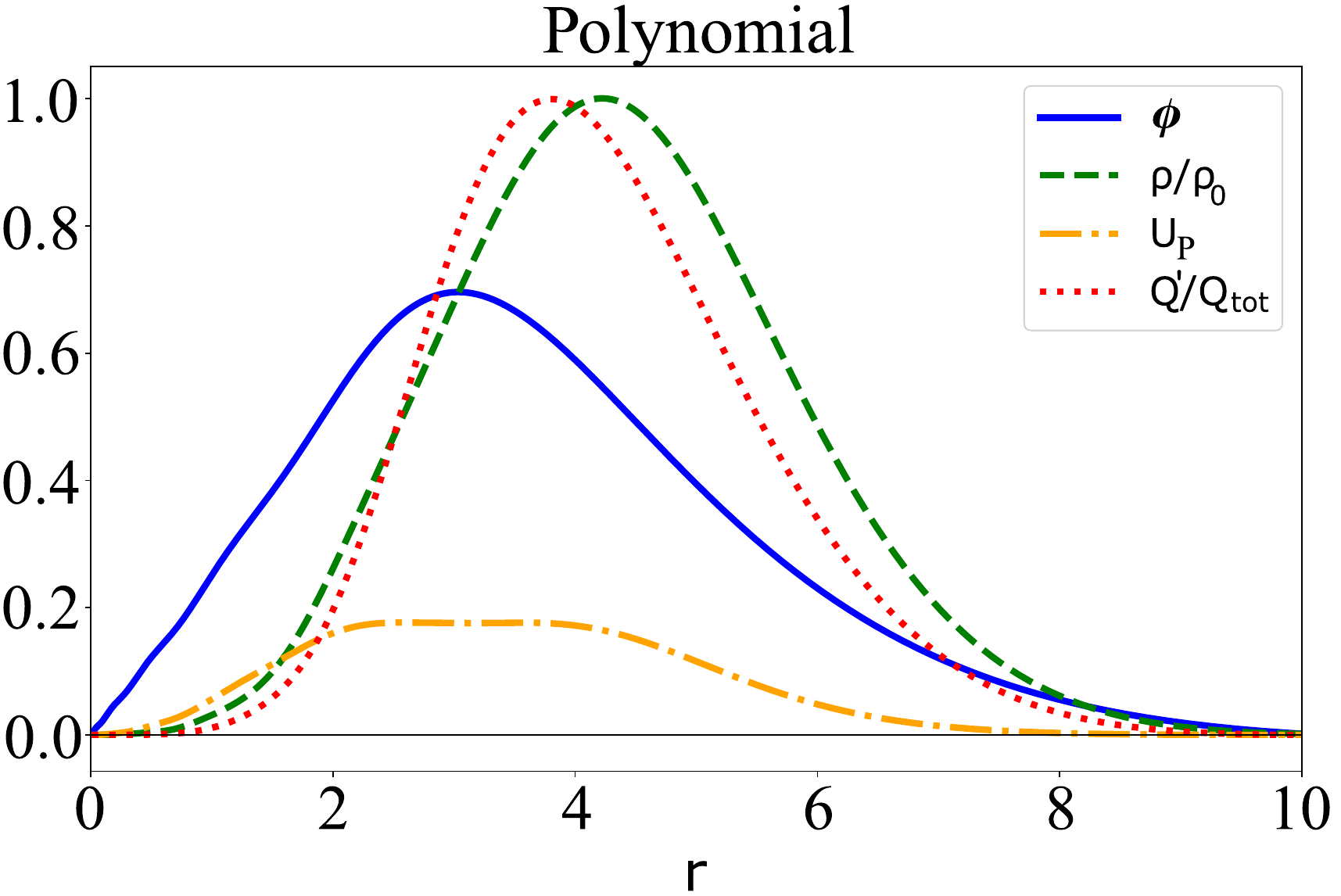}
        \end{subfigure}
        \hspace{0.0\textwidth}  
        \begin{subfigure}[b]{0.49\textwidth}
            \centering
            \includegraphics[width=\textwidth]{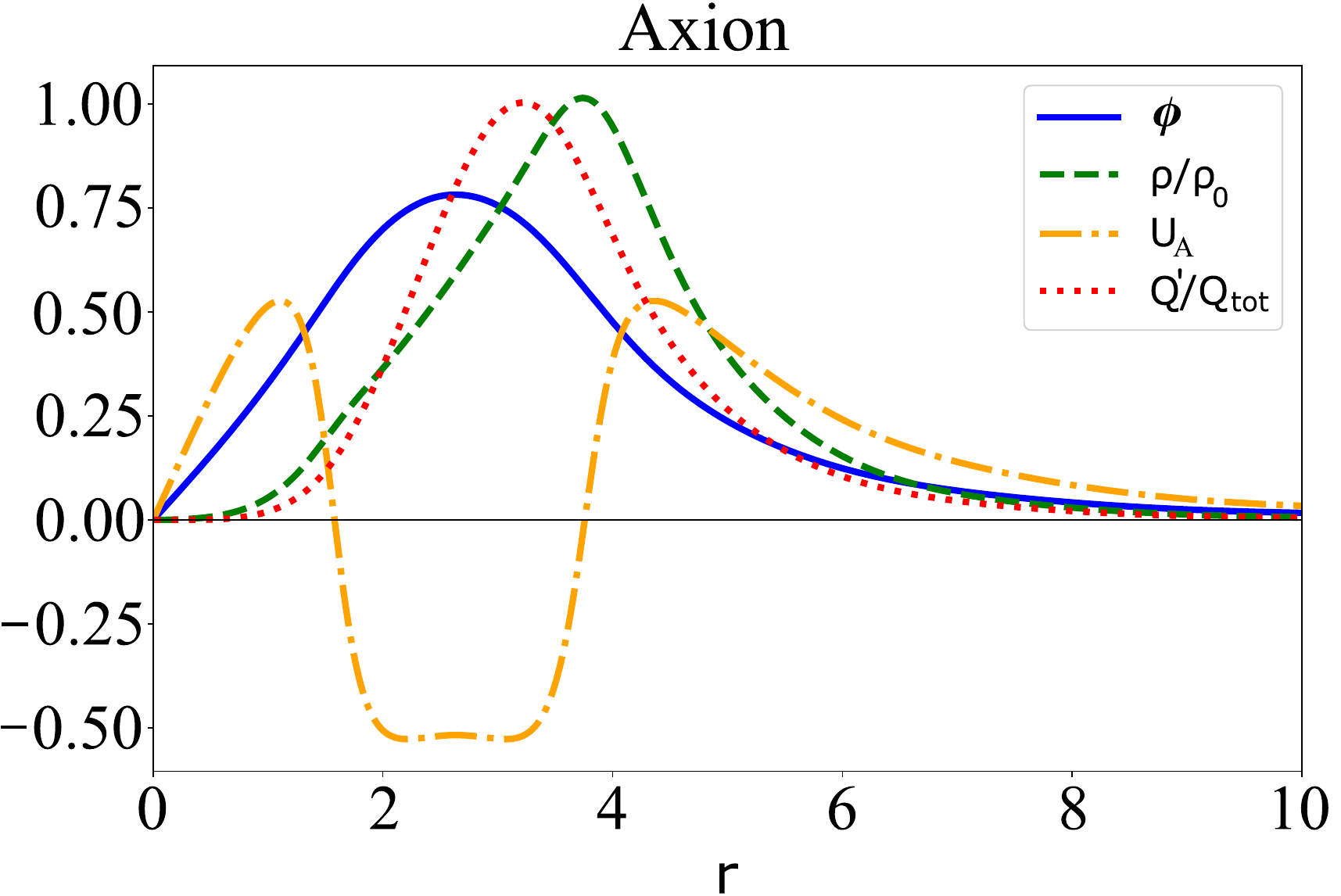}
        \end{subfigure}
    
        \caption{Radial profiles for $\omega/\mu = 0.90$ in code units ($\mu = 1$) of the scalar field $\phi(r)$ (solid blue), the normalized energy density $\rho(r)/\rho_0$ (dashed green), the differential charge density $Q'(r)/Q_{\rm tot}$ (dotted red), and the potential left: $U_P$; right: $U_A$ (dot-dashed yellow). Top: spherical ($m = 0$) solutions. Bottom: spinning ($m = 1$) solutions evaluated at the equator ($\theta = \pi/2$). Here $\rho_0$ denotes the maximum of the energy density $\rho(r)$.}
        \label{F2}
    \end{figure}
    In Fig.~\ref{F2}, we plot the radial profiles of the field ($\phi$, solid blue), potential ($U$, dot-dashed yellow), scaled Noether charge density ($Q'/Q_{\rm tot}$, dotted red) and normalised energy density ($\rho/\rho_0$, dashed green) for a representative frequency $\omega/\mu = 0.90$, for both spherical and spinning configurations of polynomial (left) and axion-like (right) $Q$-balls. In the spinning case, we show the equatorial slice $\theta= \pi/2$, where $\phi$ reaches its maximum. The field $\phi$ radial profile (solid blue) peaks differently in the spherical (top) and spinning (bottom) case. While in the spherical configuration, the maximum value is at the coordinate origin, in the spinning configuration, the regularity condition \eqref{E2.17} forces $\phi$ to be null at the centre and have its maximum at a finite non-zero radius. The energy density, $\rho/\rho_0$, follows the same pattern, giving rise to the toroidal shape. 
    
    Concerning the $Q$-ball potentials, for a spinning axion-like solution (bottom right), one can observes a region where $U_A$ becomes negative (repulsive) near the core, a feature not found in any other of the presented solutions, showing a clear behavioral difference between the polynomial and axion-like potentials even for a configuration relatively close to the $\phi \to 0$ regime -- $\omega \to \mu$, where the Taylor expansion of $U_A$ locally approximates $U_P$, \eqref{E2.14}. Additionally, in the case of a spherical configuration, while the energy density of the polynomial potential is monotonically decreasing, the solution with the axion-like potential possesses a region outside the origin where the energy density reaches a second maximum. This, while for $\omega = 0.9$ is a local maximum, passes to a global maximum for $\omega \approx 0.78$, creating an energy distribution that is no longer peaked at the centre and may possess some observational interesting signatures that we will leave for a future work.
   
\begin{figure}[h!] 
        \centering
        \includegraphics[width=1.0\textwidth]{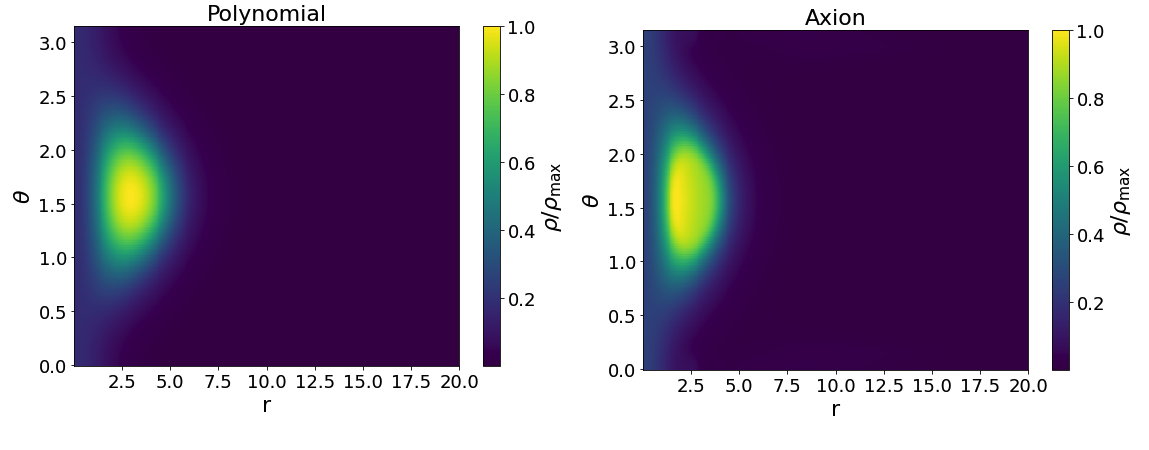} 
        \caption{2D density profiles $\rho(r, \theta)$ for $\omega/\mu=0.90$ and $m=1$, in code units ($\mu = 1$). Left: $U_P$. Right: $U_A$.}
        \label{F3}
    \end{figure}

    Fig.~\ref{F3} further shows the two-dimensional density maps $\rho(r,\theta)$ of a spinning $Q$-ball for a polynomial (left) and axion-like potential (right), at $\omega/\mu = 0.90$, normalised by $\rho_0$ so that the colour scale spans from $0$ (blue) to $1$ (yellow/white). Both maps show a single global maximum $\rho = 1$ and a smooth decline with increasing $r$. In both $U_P$ and $U_A$ we clearly see the toroidal morphology: vanishing energy density on the symmetry axis and a peak at a finite radius.
%
	\subsection{Domain of existence}	
\begin{figure}[htbp]
        \centering
    
        \begin{subfigure}[b]{0.48\textwidth}
            \centering
            \includegraphics[width=\textwidth]{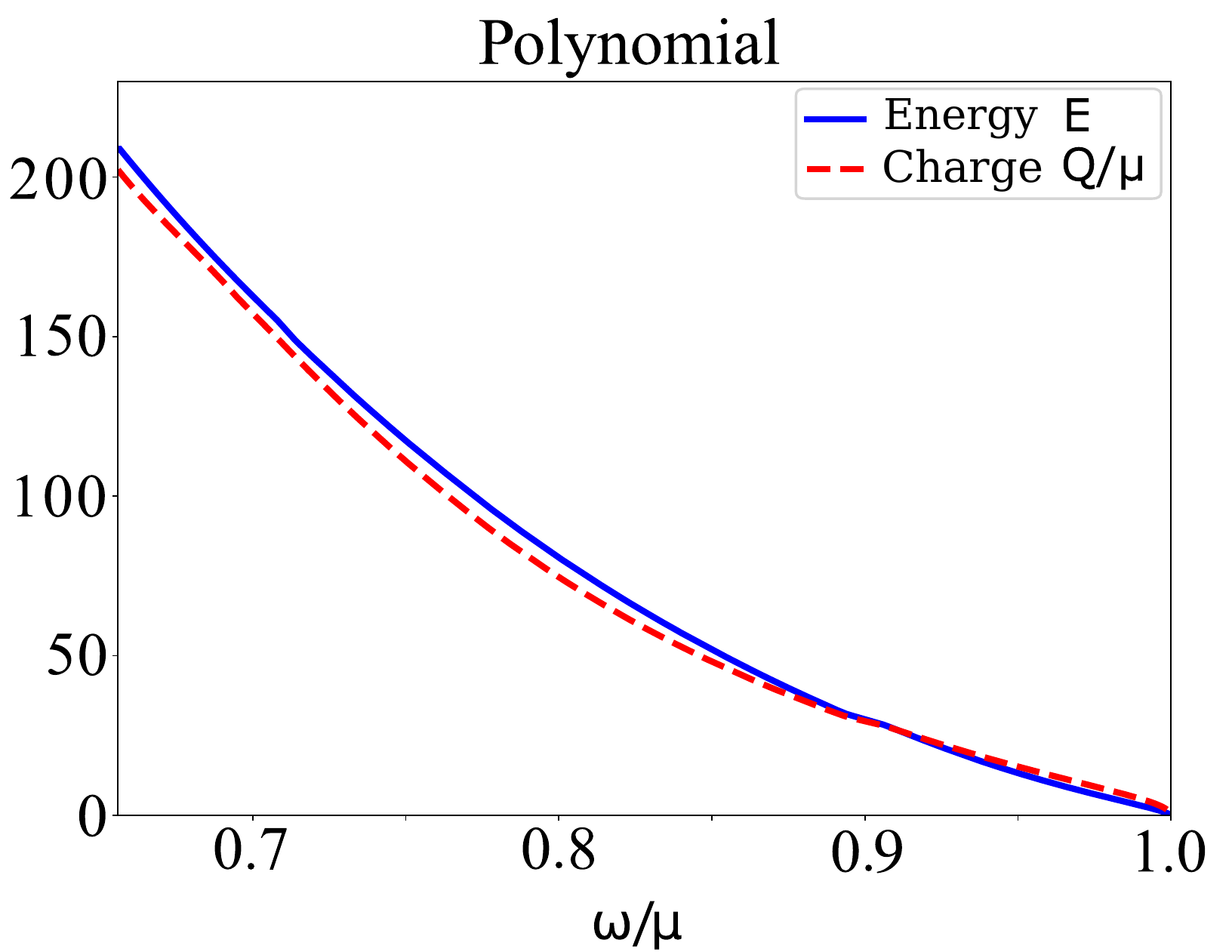}
        \end{subfigure}
        \hspace{0.02\textwidth}  
        \begin{subfigure}[b]{0.48\textwidth}
            \centering
            \includegraphics[width=\textwidth]{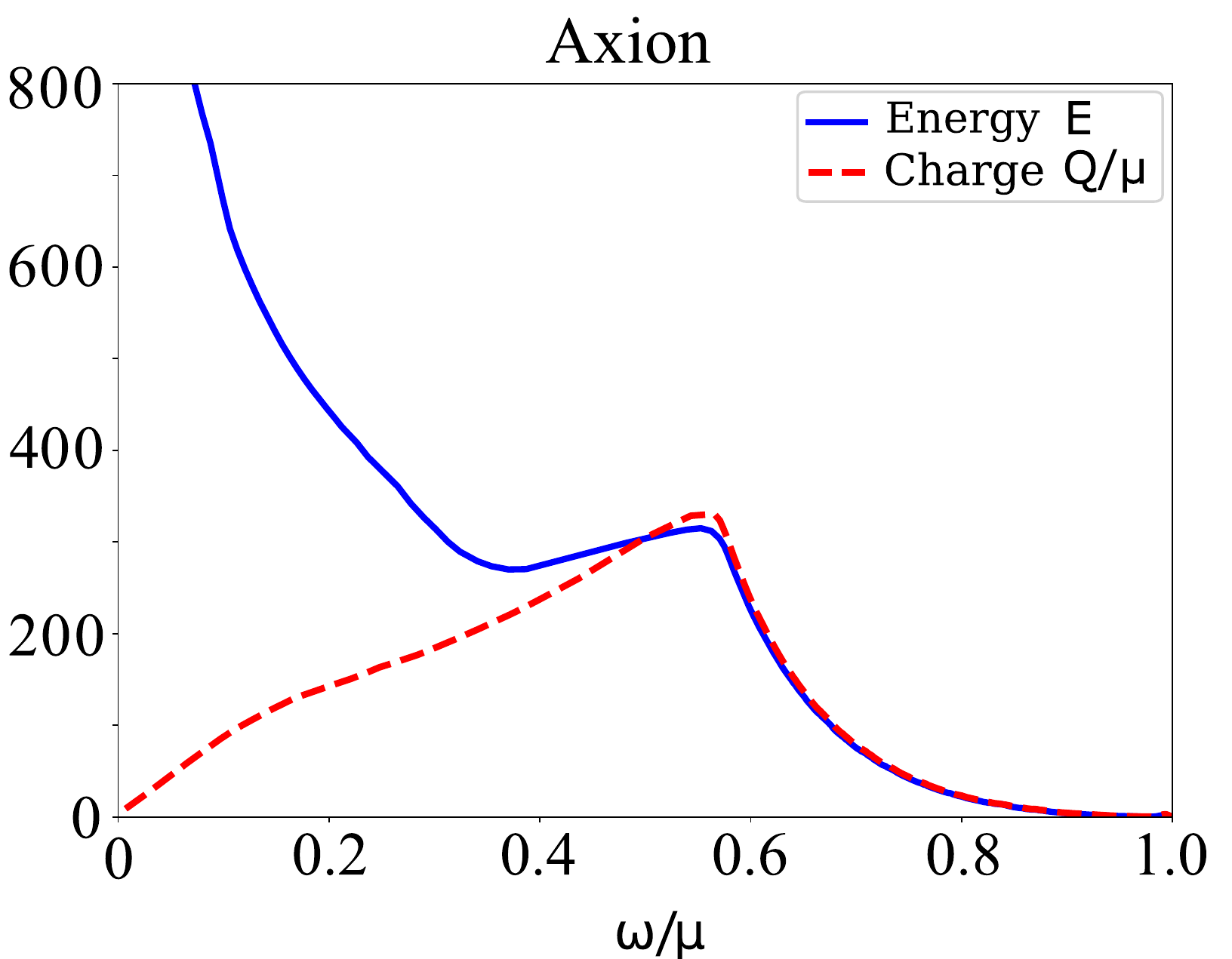}
        \end{subfigure}

        \vspace{0.5em}
        \begin{subfigure}[b]{0.48\textwidth}
            \centering
            \includegraphics[width=\textwidth]{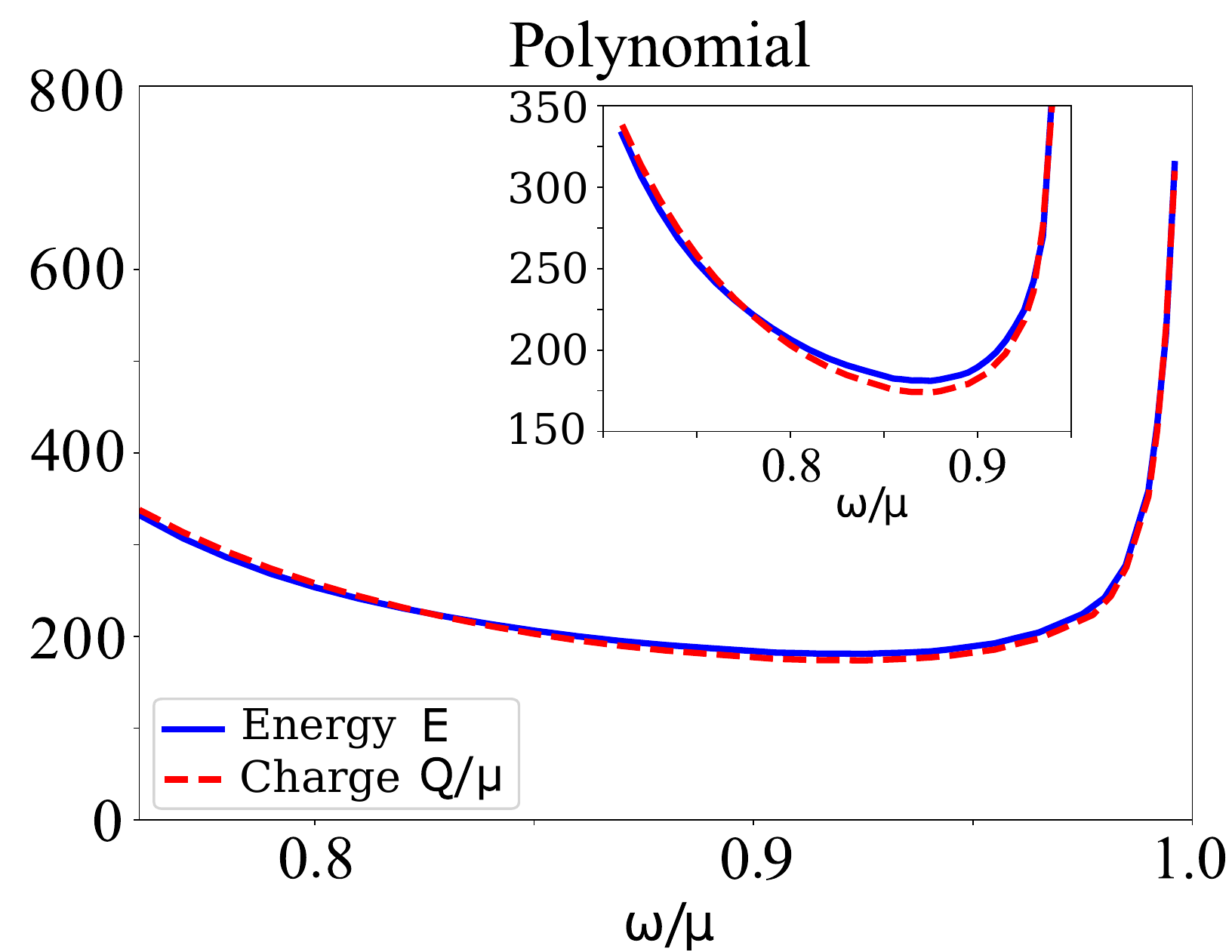}
        \end{subfigure}
        \hspace{0.02\textwidth}  
        \begin{subfigure}[b]{0.48\textwidth}
            \centering
            \includegraphics[width=\textwidth]{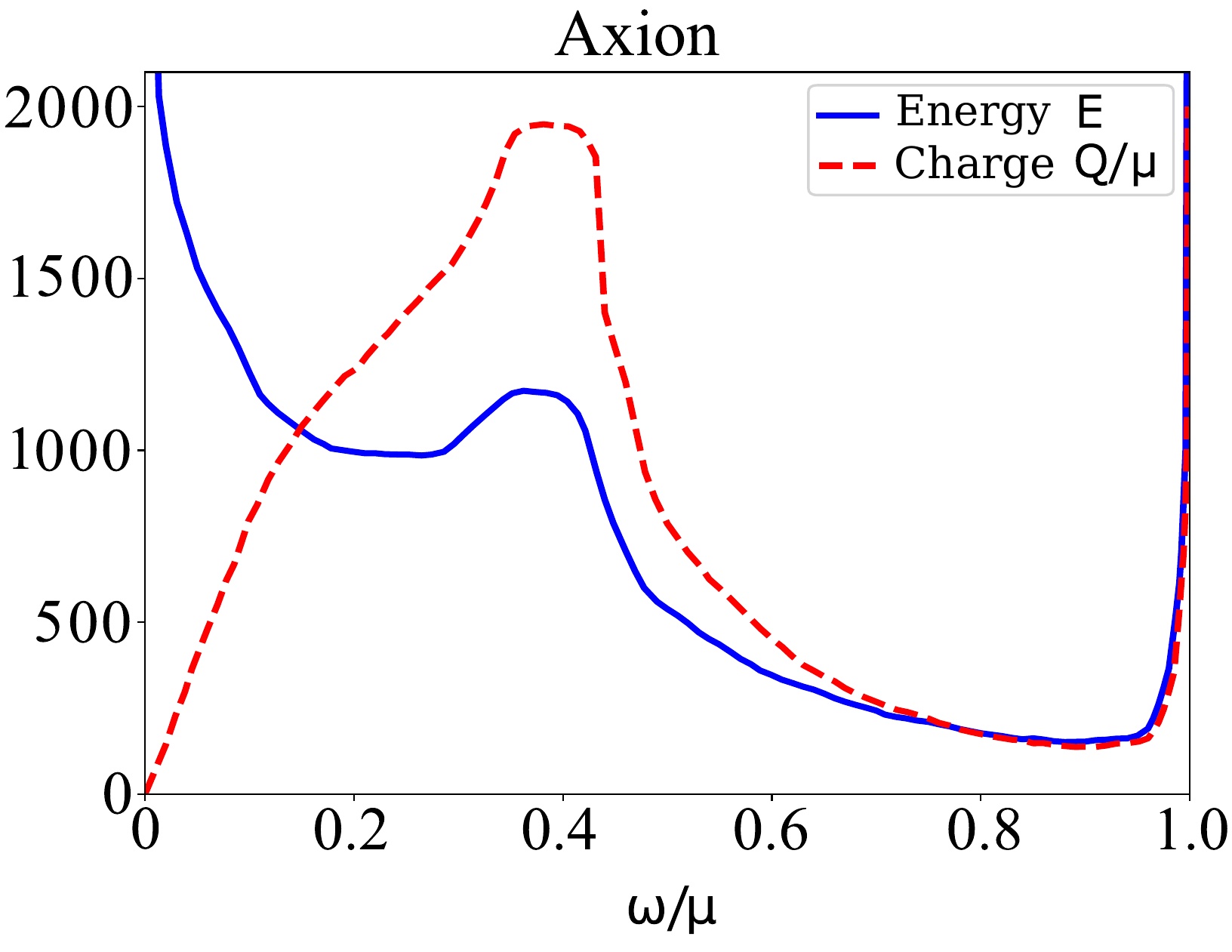}
        \end{subfigure}
    
        \caption{Domain of existence in terms of the energy, $E(\omega)$ (solid blue), and Noether charge, $Q(\omega)$ (dashed red), for: Top spherical ($m=0$); Bottom: spinning $m=1$ $Q$-ball solutions. Left: polynomial potential, $U_P$; Right: axion-like potential, $U_A$. All the quantities are in code units.}
        \label{F4}
    \end{figure}

%
    Fig.~\ref{F4} summarizes the domain of existence in terms of the energy, $E(\omega)$, and Noether charge, $Q(\omega)$, for the four configurations (spherical/spinning $\times\ U_P/U_A$). Consistent with the model discussion, the admissible frequency windows are $\omega/\mu \in [0.62, 1]$ for the spherical polynomial potential and $\omega/\mu\in [0, 1]$ for the axion-like potential. Observe that in the case of the spinning polynomial potential the frequency window follows the same behaviour as in the spherical case, demonstrating that the $Q$-balls frequency window is only mildly sensitive to
    changing $m = 0$ to $m = 1$. Near the upper edge $\omega \to \mu$, the two models coincide: for spherical solutions $E,\, Q \to 0$ (small-amplitude limit), whereas for spinning solutions $E,\, Q \to +\infty$ due to the centrifugal barrier and the growth of the ring radius.

    Before analysing the other extreme, $\omega \to \omega _-$, it is important to point out that $Q$-balls stability is typically assessed on three levels \cite{tsumagari2009physics}: \textit{i)} Classical (linear) stability, along a solution branch parametrised by the frequency $\omega$, configurations are linearly stable if $\frac{dQ}{d\omega} < 0$ -- turning points where $\frac{dQ}{d\omega}=0$ signal the onset (or loss) of one unstable mode; \textit{ii)} absolute (quantum) stability against decay into free quanta -- the state is absolutely stable if its energy per charge satisfies $E/Q < \mu$, if $E/Q > \mu$ the $Q$-ball is at best metastable to particle emission; \textit{iii)} fission (fragmentation) stability: stability against splitting into smaller $Q$-balls is ensured when the energy is sub-additive, \textit{i.e.}, $E(Q_1+Q_2) \leqslant E(Q_1)+E(Q_2)$ for all decompositions -- a practical diagnostic is that $E/Q$ decreases with $Q$ along the branch (so larger-$Q$ states do not lower energy by fission). These criteria are model-independent and apply equally to polynomial and axion-like potentials; the size of the $\omega$-window in which they are met can differ across models. 

    The behaviour as $\omega \to \omega_-$ separates the two potentials. For $U_P$ (both $m = 0$ and $m = 1$), $E$ and $Q$ grow without bound as $\omega \to \omega_-$, producing a single, monotonic branch for $m=0$ and a two branch structure (first decay and then increase) for $m=1$, see Fig.~\ref{F4} left. For $U_A$, our numerics show a qualitatively different pattern in both spherical and spinning cases: starting from $\omega \sim \mu$, $E(\omega)$ and $Q(\omega)$ first increase -- after an initial sharp decrease for $m = 1$ where a shallow minimum exists --, then reach a local maximum at $\omega \approx 0.56$ for the spherical and $\omega \approx 0.42$ for the spinning configuration; beyond that point $Q$ decreases toward zero while $E$ rises again. This creates a high-energy, low-charge continuation where $E/Q > \mu$, \textit{i.e.}, an absolutely unstable secondary branch. The location of the local extremum also partitions classical stability: the segment with $\frac{dQ}{d\omega} < 0$ (from $\omega \approx \mu$ down to the $Q$-maximum) is linearly stable, whereas the post-maximum segment with $\frac{dQ}{d\omega} > 0$ is linearly unstable.
    
    For absolute stability, the relevant band is where the curve $E/Q$ dips below $\mu$. On the axion-like branch this band is bounded by two crossings: spherical at $\omega_{\rm eq}^- \approx 0.44$ and $\omega _{\rm eq}^+ \approx 0.998$, and spinning at $\omega_{\rm eq}^- \approx 0.18$ and $\omega _{\rm eq}^+ \approx 0.78$; inside this interval $E/Q < \mu$ and the configurations are absolutely stable. The remaining branches show a narrower band or none at all (see Fig.~\ref{F4}, where the $E/Q = \mu = 1$); fission stability tracks the same trend since $E/Q$ decreases with $Q$ over the stable segments.

    Of interest for our analysis is then the region of absolute stability for both the polynomial and axion-like $Q$-ball solutions. Which we summarize to be $\omega_{\rm stab} \in [0.62,0.82]$ for the polynomial, while for the axion-like is $\omega _{\rm stab} \in [0.42,\, 0.78]$.

    Let us briefly comment on their spatial distribution. As in the spherically symmetric case, spinning $Q$-balls become larger as $\omega$ tends to the limits of the allowed interval. For $\omega \to \omega_−$, they can be viewed as squashed spheroids, homogeneously filled inside. $\phi$ increases as we move away from the central axis, reaching a maximum at the surface of the spheroid, after which it rapidly goes to zero. Solutions with $\omega \to \omega_+$, also tend to large spheroids. This time, however, they are hollow and possess the maximum of the energy density at the surface (which increases in radius as $\omega \to \omega_+$ -- being close to zero everywhere else).

    At last, we should also point out that, by \eqref{eq:angular_momentum}, the existence of a minimum value of $Q$ at $\omega_{\rm min} \approx 0.92$ for the $m=1$ polynomial solution creates a restriction on the total angular momentum, $J$. Polynomial $Q$-balls are not allowed to rotate arbitrarily slowly; there is a discrete spectrum of spinning excitations. On the other hand, for an axion-like potential, while the local minimum value of $Q$, still exists, there is a branch which connects directly with the $Q=0$ solution and hence no minimum value of angular momentum exists for this solutions. 

 \section{Building rotation curves of a galaxy-size $Q$-ball}\label{S6}
%
    In order to derive the expected rotation curve generated by a $Q$-ball configuration, we will consider its physical energy density, $\varepsilon$, as a source of a gravitational potential $\psi(r,\theta)$ in the Newtonian limit.
    Although the density of a $Q$-ball configuration is derived under the assumption that gravitational effects are negligible compared to the field’s self-interaction, the total mass of the configuration can, nonetheless, produce a gravitational potential comparable to that of a realistic galaxy-sized DM halo, influencing the motion of visible matter. From \eqref{E2.7}, we have
    \begin{equation}\label{E6.1}
       \rho (r,\theta)\equiv \varepsilon(r,\theta) =  \omega ^2 \phi ^2 +(\partial_r \phi)^2+\frac{(\partial_\theta \phi)^2}{r^2}+\frac{m^2 \phi ^2}{r^2 \sin ^2 \theta}+U(\phi)\ .
    \end{equation}
    In natural units ($\hbar=c=1$), the energy density $\varepsilon(R,z)$ and the mass density $\rho(R,z)$ coincide.
    
    When converting to astrophysical units, one simply reinstates the appropriate factor of $c^2$ through $\rho = \varepsilon / c^2$.
    When studying $Q$-balls as disk galaxies' DM halos, we are mainly focused on quantities defined at the equatorial plane -- where observational data are obtained. As such, it is convenient to switch from spherical coordinates $(r,\theta,\varphi)$ -- where $Q$-balls are numerically computed --, to cylindrical coordinates $(R,z,\varphi)$. This transformation enables a direct analysis of galactic rotation curves based on the gravitational potential and velocity profiles in the $z=0$ plane. In cylindrical coordinates $(R,z,\varphi)$, the field ansatz becomes:
    \begin{equation}
     \Phi(R,z,\varphi,t) = {\phi}(R,z)\, e^{i(m \varphi-\omega t)}\ , \qquad z = r \cos\theta\ ,
    \end{equation}
    and the energy density reads:
    \begin{equation}
     \varepsilon(R,z) = \omega^2 {\phi}^2  + \big(\partial_R {\phi}\big)^2 + \big(\partial_z {\phi}\big)^2 + \frac{m^2}{R^2}{\phi}^2 + U({\phi})\ .
     \label{eq:energy_density}
    \end{equation}
    So far, we have expressed the equations in dimensionless, ``code'', units ($\mu = 1$). By re-introducing all the physical constants, the code units are connected to the physical ones via the radial mass scale $\mu = m_\phi/(\hbar c)$, with $ m_\phi$ the particle mass expressed in electron-volts (eV):
    \begin{equation}\label{E6.4}
     \hat{R} = \mu R\ ,\qquad \hat{z} = \mu z\ ,\qquad \hat{\rho}(\hat{R},\hat{z}) = \omega^2 \hat{\phi}^2 + \big(\partial_{\hat{R}} \hat{\phi}\big)^2 + \big(\partial_{\hat{z}} \hat{\phi}\big)^2 + \frac{m^2}{\hat{R}^2}\hat{\phi}^2 + U(\hat{\phi})\ ,
    \end{equation}
    where now we have used  $\hat{X}$ to explicitly highlight a quantity written in code units. For a given frequency, $\omega/\mu$, the corresponding physical mass density, acting as the source term of the Newtonian gravitational potential, is recovered through:
    \begin{equation}\label{E6.5}
     \rho(R,z;\mu,\rho_0) = \rho_0\, \hat{\rho}(\mu R,\mu z)\ ,
    \end{equation}
    where $\rho_0$ carries the units of mass density. The gravitational potential, $\psi(R,z)$, is then obtained from the mass density through the Poisson equation:
    \begin{equation}\label{eq:poisson}
     \nabla^2 \psi(R,z;\mu,\rho_0) = 4\pi G\, \rho(R,z;\mu,\rho_0)\ .
    \end{equation}
    At the equatorial plane ($z=0$), the circular velocity associated with the $Q$-ball energy distribution is given by:
    \begin{equation}\label{eq: circular_vel}
     v_Q^2(R;\mu,\rho_0) = R\, \frac{\partial \psi}{\partial R}(R,0;\mu,\rho_0)\ .
    \end{equation}
    The quantities $(\mu, \rho_0)$ are determined by matching the dimensionless solution to observational data (see Section \ref{S7}). Specifically, $\mu$ sets the radial scale, while $\rho_0$ is proportional to the total halo ($Q$-ball) mass. From the solution of \eqref{eq:poisson}, the gravitational potential in the equatorial plane generated by an axially symmetric mass distribution is given by ~\cite{BinneyTremaine}:
    \begin{equation}\label{E6.8}
     \psi(R,0) = -4G \int_{0}^{\infty} \text{d}R' \int_{-\infty}^{\infty} \text{d}z'\;\frac{R'\,\rho(R',z')}{\sqrt{(R+R')^2+z'^2}}\;K\!\left(\frac{2\sqrt{RR'}}{\sqrt{(R+R')^2+z'^2}}\right)\ ,
    \end{equation}
    where $K(k)$ is the complete elliptic integral of the first kind and $\rho(R',z')$ is the physical mass density. 
    Substituting the correct scaling \eqref{E6.4} from physical to code units for a $Q$-ball, into the integral \eqref{E6.8}, one obtains the dimensional factorization: 
%
    \begin{equation} \label{E6.9}
     \psi(R,0;\mu,\rho_0) \;=\; -\frac{4G\rho_0}{\mu^2}\,\hat{\psi}(\hat{R})\ ,\qquad \hat{\psi}(\hat{R})\propto\int \text{d}\hat{R}'\,\text{d}\hat{z}'\;\hat{R}  R'\,\hat{\rho}(\hat{R}',\hat{z}')\ \ldots
    \end{equation}
    and therefore the circular velocity separates into a dimensional pre-factor times a dimensionless, model-dependent curve:
    \begin{equation}\label{E6.10}
      v_{Q}^2(R;\mu,\rho_0) \;=\; \frac{4G\rho_0}{\mu^2}\;v_{\rm code}^2(\hat{R})\ , \qquad v_{\rm code}^2(\hat{R})\equiv -\hat{R}\frac{\text{d}\hat{\psi}}{\text{d}\hat{R}}\ ,
    \end{equation}
    with $\hat{R}=\mu R$. The mass density amplitude, $\rho_0$, can be further rewritten in terms of the total halo mass. Defining the dimensionless mass (energy) integral:
    \begin{equation}
     {\rm E} \;=\; 2\pi\int_0^\infty \text{d}\hat{R}\int_{-\infty}^{\infty} \text{d}\hat{z}\; \hat{R}\,\hat{\rho}(\hat{R},\hat{z})\ ,
    \end{equation}
    the physical halo mass $M_{\rm h}$ and $\rho_0$ are related by $M_{\rm h}=\rho_0\,{\rm E}/\mu^3$, hence $\rho_0 = M_{\rm h}\,\mu^3/{\rm E}$. Replacing this into \eqref{E6.10} yields the simple form
    \begin{equation}\label{E6.12}
     v_{Q}^2(R;\mu,M_{\rm h},\omega) \;=\; \frac{4G\,M_{\rm h}\,\mu}{{\rm E}}\; v_{\rm code}^2(\mu R)\ ,
    \end{equation}
    which depends only on measurable or computable quantities: the halo mass $M_{\rm h}$, the radial scale $\mu$ (related to the particle mass), the dimensionless energy integral ${\rm E}$ (determined by the $Q$-ball profile, Section~\ref{S4}) and the dimensionless rotation curve $v_{\rm code}(\hat{R})$ computed from the field solution. Equations \eqref{E6.9}-\eqref{E6.12} are the operational relations used to fit $Q$-ball models to observed rotation curves.
    
    The shape of $v_{\rm code}(\hat{R})$ is set exclusively by the dimensionless density profile $\hat{\rho}(\hat{R},\hat{z})$, and therefore by the chosen self-interaction potential, $U$. In the polynomial case, the large $\phi$ growth of $U_P\propto\phi^6$ strongly penalises extended, high-amplitude field configurations, producing more compact, strongly localised $Q$-balls. By contrast, an axion-like potential can support broader field profiles that explore non-linear plateaus of $U_A$. Consequently, when translated to physical units via \eqref{E6.12}, polynomial $Q$-balls tend to give rotation curves associated with relatively concentrated halos for a given particle mass $m_\phi$, whereas axion-like $Q$-balls can produce more extended halo contributions for the same dimensionless solution family -- a qualitative distinction that must be confronted with data when fitting observed rotation curves.
%

%
\section{Data}\label{S5}
%
    In this work we use the \textsc{SPARC} (\textit{Spitzer Photometry \& Accurate Rotation Curves}) database\footnote{\href{https://astroweb.cwru.edu/SPARC/?_gl=1*1t6sicr*_gcl_au*MjQwMTE4MTU4LjE3NTEwMTcwMzU.}{https://astroweb.cwru.edu/SPARC/}} \cite{lelli_2016_sparc}, which provides homogeneous 3.6 $\mu$m photometry and accurate rotation curves for 175 nearby disk galaxies. 

    The sample spans a wide range of morphologies (S0--Im), luminosities ($10^{7}\!<\!L_{[3.6]}/L_\odot\!<\!10^{12}$), and circular velocities ($20$–$300$\,km\,s$^{-1}$), encompassing the diversity of spiral and irregular galaxies in the local Universe.

    The rotation curves were derived from interferometric H\textsc{i} and optical H$\alpha$ observations collected over the past three decades with different facilities, the Westerbork Synthesis Radio Telescope (WSRT), the Very Large Array (VLA), the Australia Telescope Compact Array (ATCA), and the Giant Metrewave Radio Telescope (GMRT). Tilted-ring modelling was applied to the velocity fields, with quality flags (Q $=4-3$) indicating the reliability of each curve; typical velocity uncertainties are a few km\,s$^{-1}$. Distances were obtained from standard candles when available, from cluster membership (\textit{e.g.} Ursa Major at $18\pm0.9$\,Mpc), or from Hubble-flow estimates corrected for Virgo-centric infall, assuming $H_0=73$\,km\,s$^{-1}$\,Mpc$^{-1}$.

    The 3.6\,$\mu$m surface photometry was extracted from \textit{Spitzer}/IRAC imaging, primarily from the Spitzer Survey for Stellar Structure in Galaxies, S$^4$G \cite{sheth_2010}, and processed with the \textsc{Archangel} pipeline \cite{schombert_2011}. Exponential fits to the outer profiles yield disk scale lengths, while bulge components (for $T<4$) are modelled as spherical. The gaseous contribution, $V_{\rm gas}$, is computed from the H\textsc{i} surface-density profiles following \cite{casertano_1983}, accounting for the contribution of helium. The total baryonic contribution to the rotation curve is then
    \begin{equation}\label{E5.1}
     V_{\rm bar} (R; \Upsilon_*)\equiv  V_{\rm bar}^2 = \Upsilon_{\rm disk} V_{\rm disk}|V_{\rm disk}| + \Upsilon_{\rm bul} V_{\rm bul}|V_{\rm bul}|+ V_{\rm gas}|V_{\rm gas}|\ ,
    \end{equation}
    as in Eq.~(2) of \cite{lelli_2016_sparc}, where they included the absolute values to reflect the fact that the gas contribution to $V_{\rm bar}^2$ can sometimes be negative. Above, $\Upsilon_{\rm disk} \equiv \Upsilon_{*}$ and $\Upsilon_{\rm bul} \equiv 1.4\Upsilon_{*}$ are the stellar mass-to-light ratios for the disk and the bulge, respectively. In \cite{lelli_2016_sparc} the mass-to-light ratios at 3.6\,$\mu$m are fixed to $\Upsilon_*=0.5\,M_\odot/L_\odot$ and $\Upsilon_\text{bul} = 0.7\,M_\odot/L_\odot$, consistent with stellar population synthesis models using standard initial mass functions (IMFs) \cite{schombert_2014a, mcgaugh_2014}. These values minimise the scatter in the baryonic Tully–Fisher relation and yield realistic gas fractions and degrees of baryonic maximality, while lower $\Upsilon_\star$ values lead to unphysically sub-maximal disks \cite{lelli_2016_sparc}. In the following, we will account for the variation in $\Upsilon_*$ for each galaxy by optimising it within the fit, assuming an informative Gaussian prior centred on 0.5, $\mathcal{N}(0.5,\sigma=0.1)$, we computed the corresponding $\Upsilon_\text{bul}=1.4\,\Upsilon_*$. Note that here, we implement a simple proof-of-concept application of our $Q$-ball solutions; refined analyses will be applied in a subsequent paper for more sophisticated bosonic solitonic models. Each SPARC entry provides the observed rotation curve and the modelled contributions of gas, stellar disk, and bulge, forming one of the most comprehensive datasets for galaxy mass modelling and DM studies.
%

    \section{Results}\label{S7}
%
    Among the 175 galaxies in the SPARC catalogue, in order to minimise selection effects, we consider disk galaxies with at least ten measured rotation-curve points, resulting in a final sample of 116 objects spanning a wide range of masses and morphologies. This choice allowed us to assess the phenomenological signature of the $Q$-ball model across a variety of disk galaxy population, rather than focusing on a restricted subsample optimised for a specific halo profile.
    
    The total rotation curve of a galaxy is then expressed by including the contribution of the $Q$-ball:
    \begin{equation}\label{E7.1}
        v_\text{tot}(R;\mu, M_{\rm h}, \omega, \Upsilon_*) = \sqrt{V_{\rm bar}^2(R; \Upsilon_*)+v_Q^2(R ;\mu, M_\text{h},\omega)}\ ,
    \end{equation}
    where $v_Q$ is defined by \eqref{E6.10}, and $V_{\rm bar}$ reflects the sum of the baryonic components as given by \eqref{E5.1}. We fit \eqref{E7.1} against RC SPARC data, assuming flat priors in $\omega \in[0.383, 0.999]$\footnote{For the polynomial $Q$-ball, the interval is tighter, $\omega > 0.62$.}, $\text{log}_{10}(m/\text{eV}) \in[-31, -23]$, and $\text{log}_{10}(M_{\rm h}/\text{M}_\odot) \in[6, 18]$. Note that, as mentioned above, the stellar mass-to-light ratio $\Upsilon_*$ is also optimised within the fit with an informative Gaussian prior centred on $\Upsilon_*=0.5$, with a standard deviation of 0.1.

    We consider both the polynomial and the axion-like potential models for the scalar field, sampling 50000 points in the parameter space with a Monte Carlo Markov-Chain (MCMC) method; we roughly compare the performances of the two models by using the Bayesian Information Criterion (BIC),
    \begin{equation}
        \text{BIC} = k\ln(n) -2\ln(\ell)\ ,
    \end{equation}
    where $k$ is the number of free parameters, $n$ the number of data points and $\ell$ the likelihood of the fit. A model $A$ is significantly preferred over a model $B$ if $\Delta \text{BIC} = \text{BIC}_B - \text{BIC}_A \gtrsim 6$. 
    Since the models have the same number of free parameters, $\Delta$BIC differs only by the fit likelihood term; under Gaussian errors this is equivalent (up to a constant) to a $\Delta \chi^2$ comparison. 

    To evaluate the viability of the $Q$-balls, we further compare both the polynomial and the axion-like models with a fit obtained assuming standard Navarro-Frenk-White (NFW) halo profile \cite{Navarro96} for the DM distribution. The latter's associated circular velocity at radius $r$ reads:
    \begin{equation}
        v^2_{NFW}(r) = \frac{G M_{NFW}(r;r_{200},r_s) }{r}\ ,
    \end{equation}
    where 
    \begin{equation}
        M(r;r_{200},r_s)  = M_{200} \frac{\ln(1 + r/r_s) - (r/r_s)/(1+r/r_s)}{\ln(1+c_{200}) - c_{200}/(1+c_{200}) }\ .
    \end{equation}
    With $r_{s}$ and $r_{200}$ are the two free parameters characterising the halo, $c_{200} = r_{200}/r_s$ and $M_{200} = (100  H^2(z)/G)\times r_{200}^3 $. Finally, $H(z)$ is the Hubble parameter at redshift $z$. 
     
\begin{figure}[h!]
	\centering
	\includegraphics[width=1.0\textwidth]{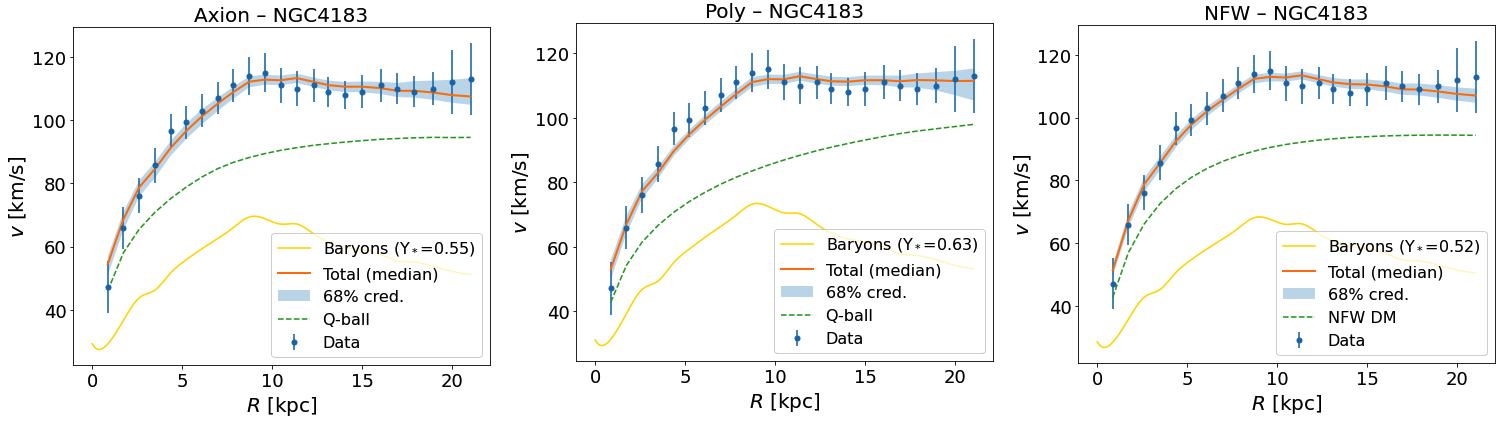}\\
	\caption{Example of fitted rotation curves for the galaxy NGC 4183 of the SPARC sample. The orange solid lines and the light blue region indicate the best fit model from \eqref{E7.1} and the 68\% confidence region. Blue points with error bars are the observed rotation curves. The green dashed lines represent the best-fit rotation curve contributions of the $Q$-ball (left and central plots) and of the NFW DM profile (right), while the yellow lines refer to the combined contribution of the baryonic components. Left: axion-like potential, $U_A$. Central: polynomial potential, $U_P$. Right: standard NFW DM distribution.}
	\label{fig:galaxyExample}
    \end{figure}

     In Fig.~\ref{fig:galaxyExample} we show, as an example, the best fit axion-like and polynomial $Q$-ball rotation curves (green dashed lines in the left and central plots), the baryonic contribution (yellow curves) and the total contribution (baryonic + $Q$-ball) along with its 68\% confidence region (orange solid lines and light blue shaded area) compared to the observed curve (blue points with error bars) for the spiral galaxy NGC 4183 in the Ursa Major cloud.  
    For the chosen galaxy, the $Q$-ball model fits reasonably well the observed rotation curve, with a reduced $\chi^2_{\rm red} = 0.35,\ 0.26$ for the polynomial and axion-like case, respectively. The axion-like potential performs slightly better ($\Delta$ BIC $= 1.68$) against the polynomial potential, which is a common trend found on all the galaxies in our sample. When compared against the standard NFW profile, Fig.~\ref{fig:galaxyExample} (right), the axion-like $Q$-ball is less preferable, with a difference of $\Delta $BIC$ = 3.02$ ($\chi^2_{\rm red} = 0.18$) in favour of the NFW profile.
    
\begin{figure}[h!]
	\centering
	\includegraphics[width=1.0\textwidth]{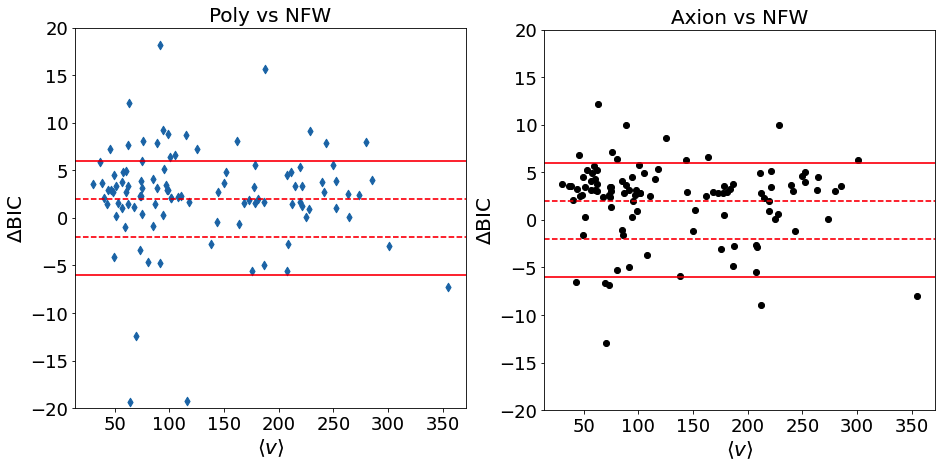}\\
	\caption{Distribution of $\Delta$BIC = $\text{BIC}_{P/A} - \text{BIC}_{NFW}$ of the polynomial and axion-like models with respect to the NFW DM profile (a $\Delta$BIC $> 0$ favours NFW), as a function of the average (positive) circular velocity of the galaxy $\langle v \rangle$ The dashed and red horizontal lines represent the limits $\Delta$BIC = $\pm2$ and $\pm 6$, marking the regions above which the preference for a given model becomes mild and strong, respectively. Left: Polynomial - NFW. Right: Axion - NFW.}
	\label{fig:bic}
    \end{figure}

    Overall, the NFW model provides the best description of the majority of rotation curves in the sample. However, the $\Delta$BIC distribution, displayed in Fig. \ref{fig:bic}, shows that the $Q$-ball models very often yield statistically reasonable fits, with $\Delta$BIC values typically in the range 2–5, and outperforms NFW in a non-negligible subset of galaxies.

    It is important to remark here that we are adopting an illustrative, simplified model which neglects the gravitational term in the scalar field action and does not account for the complex physical mechanisms in the galactic environment. Thus, this comparison should be considered cautiously as a proof-of-concept for future astrophysical applications. However, the toroidal morphology characteristic of spinning $Q$-balls seems to translate, in the equatorial plane, into a distinctive family of circular–velocity profiles $v(R)$ that can reproduce the dark component of disk–galaxy rotation curves.

\begin{figure}[h!]
	\centering
	\includegraphics[width=1.0\textwidth]{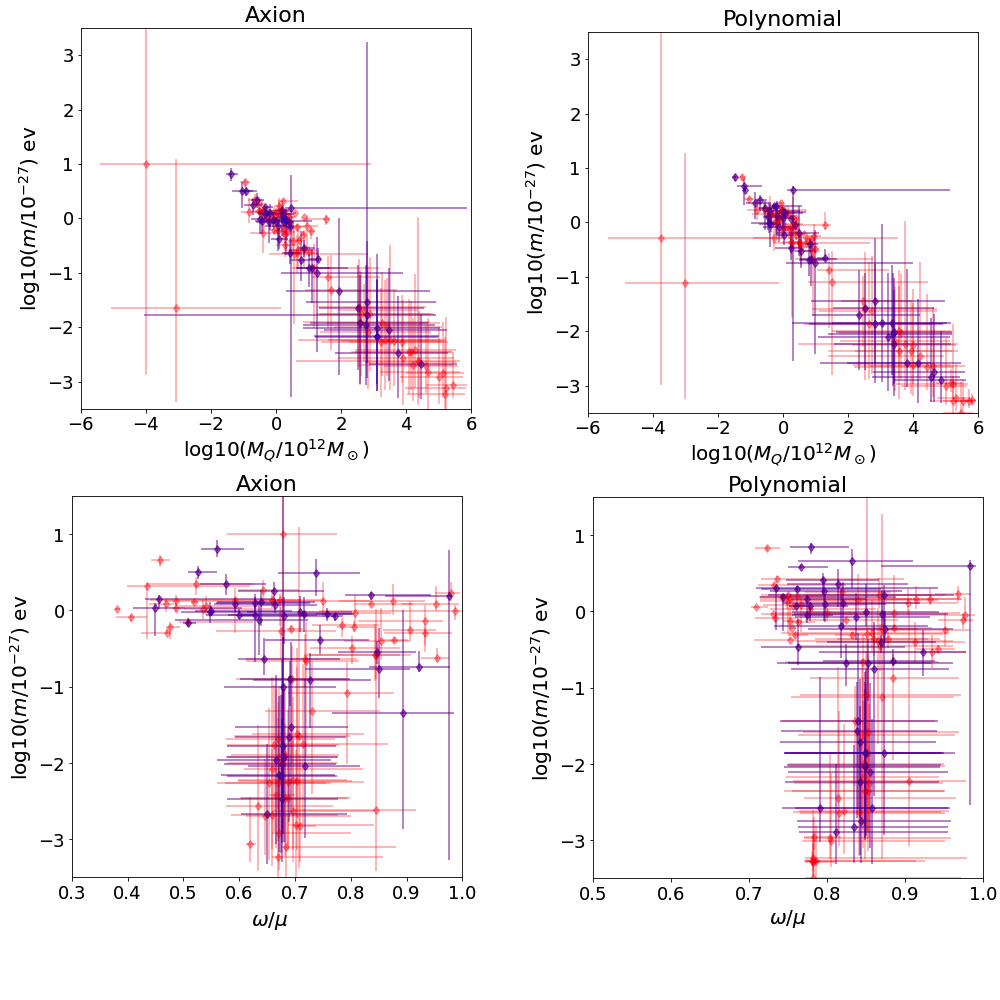}\\
	\caption{Comparison of the best fit parameters with their 68\% confidence interval for the axion-like (left) and polynomial (right) $Q$-ball models. Top: $\text{log}_{10}(M_{\rm h}/\text{M}_\odot)$ \textit{vs.} $\text{log}_{10}(m/\text{eV})$. Bottom: $\omega$ \textit{vs.} $\text{log}_{10}(m/\text{eV})$. In all plots, the red points indicate a $(\chi^2_{\rm red})^X > 1.0$, where $X = P, A$ for polynomial and axion-like, respectively.}
	\label{fig:resultmass}
    \end{figure}

    Fig.~\ref{fig:resultmass} shows the estimated best fit log particle mass, log halo mass and frequency $\omega$ parameters with their 68\% intervals. The red points and bars indicate fits with a $\chi^2_{\rm red}$ larger than 1.0. For both polynomial and axion-like potentials, we found that in the majority of the cases (39 galaxies) the particle mass is constrained to be between $- 28 < \log_{10} (m/\text{eV}) < - 26.5$, which is consistent with the results presented in \cite{Mourelle24,Mourelle25} and the expected maximum mass for ultra-light DM models \cite{Marsh_2022,lopez2025scaling}.
    However, for a few objects, the best fit mass $m$ reaches quite small values, down to $10^{-30}$ eV, with a corresponding halo mass of $\log_{10} (M_\text{h}/\text{M}_\odot) \sim 16-18 $; these values are unrealistic for a galactic-scale $Q$-ball, corresponding of a physical size of the halo of several Mpc, much larger than a cluster of galaxy. This is not surprising as our simplified model is not able to fully capture the wide variety of galaxy environments, morphologies and dynamics/evolutions, as well as the simultaneous gravitational back-reaction between the soliton and the galaxy, as mentioned above. Interestingly, the galaxies showing this unphysical behaviour are localised in an interval of frequencies around 0.7 and 0.8 for the axion-like case. 

   The average constraints on the free parameters accounting for the scatter among the single galaxy estimates are given by
    \begin{equation}
        \text{log}_{10}(M_{\rm h}/\text{M}_\odot) = 12.65 \pm 0.07 \,, \qquad \text{log}_{10}(m/\text{eV})= - 27.35 \pm 0.04 \,,\qquad \omega = 0.81 \pm 0.01\,,
    \end{equation}
    for the polynomial potential, and
    \begin{equation}
        \text{log}_{10}(M_{\rm h}/\text{M}_\odot) = 12.44 \pm 0.06 \,, \qquad \text{log}_{10}(m/\text{eV})= - 27.15 \pm 0.04 \,,\qquad \omega = 0.69 \pm 0.02\,,
    \end{equation}
    for the axion-like case. Both cases inside the stability region of the corresponding $Q$-ball solutions: polynomial, $\omega_{\rm stab} \in [0.62,\,0.82]$; and axion-like $\omega _{\rm stab} \in [0.42,\, 0.78]$.     
%
\section{Conclusions}\label{S8}
%
    In this paper, we examined whether dynamically robust, rotating scalar $Q$-balls can reproduce the DM contribution to disk-galaxy rotation curves. We presented a theoretical and numerical framework for two classes of self-interactions: the standard sextic polynomial potential, $U_P$, and an axion-like periodic potential, $U_A$, that reduces to the polynomial form at small amplitudes ($\omega \to \mu)$, but departs from it at high (non-linear regime).
    
    To compute these solutions efficiently and in an approximated closed form, we presented for the first time the \texttt{SpectralPINN}, an in-house developed hybrid solver that embeds pseudo-spectral basis functions directly inside a PINN optimisation loop. The result is a solver that preserves the exponential accuracy of spectral approaches while retaining the flexibility of neural networks residual minimisation and yields compact representations of both spherically symmetric and spinning $Q$-balls for the two potentials considered. 
    
    The computed solutions, and their domain of existence, are analysed for the first time in the axion-like case. For the spinning and spherical configurations, both models coincide at $\omega \to \mu$ as expected from the low field approximation. On the other hand, toward $\omega \to \omega_-$ the polynomial $U_P$ is monotonically increasing (after reaching a minimum for the spinning case) ($E,\,Q\to + \infty$), while the axion-like solution reaches a local (global) maximum of $E$ ($Q$) (once again after reaching a minimum for the spinning case), turns over and continues to a high-energy/low-$Q$ branch with $E/Q>\mu$. Absolute stability ($E/Q<\mu$) occurs only in an intermediate band -- narrow for $U_P$, broader for $U_A$ -- while classical stability tracks segments with $dQ/d\omega<0$. 

    An empirical assessment on a quality-controlled subset of 116 galaxies from the SPARC catalog \cite{lelli_2016_sparc} shows that both potentials can fit observed rotation curves with statistically acceptable quality in many systems, and can be competitive with the standard NFW DM model in a subset of galaxies. More in detail, the axion-like potential is mildly preferred over the polynomial, delivering the lowest BIC in 22 systems, while the polynomial is preferred only for 10 galaxies. While the NFW provides the lowest BIC for 84 systems, the preference is strong only for a small ($\lesssim 10\%$) fraction of the sample.
    
    In most galaxies the inferred scalar-particle mass clusters around $m_\phi\sim10^{-27}\,\mathrm{eV}$, in line with expectations for solitonic DM scenarios on galactic scales (\textit{e.g.} \cite{Marsh_2022,Mourelle24}). Around 30\% of the objects analysed favour unrealistically low masses and excessively large halo normalisations; however, these can be considered as artifacts of our simplified setup and of residual systematics in individual rotation curves, rather than viable astrophysical solutions. 

    We emphasize that the present work is intentionally proof-of-concept. By construction, we neglect the scalar field self–gravity in the equations of motion and adopt a simplified description of the galactic environment, which does not account for all observational systematics. These choices isolate the core question of phenomenological viability, but naturally limit the completeness of the inference and imply that we do not expect to outperform empirical halo profiles such as NFW across the full galaxy sample.

    The natural prosecution of this analysis will incorporate self-gravity directly in the action and confront the model jointly with complementary probes, including stellar and gas kinematics, gravitational lensing, and the baryonic Tully-Fisher relation. Moreover, extending the sample to include dwarf-galaxy rotation curves may provide additional insight into the ultra-light and fuzzy DM regime, with several studies suggesting a larger particle mass, $m \gtrsim 10^{-23},\mathrm{eV}$ (\textit{e.g.}~\cite{Zoutendijk21,Banares23}).

    Taken together, our results identify rotating scalar $Q$-balls, especially with axion-like self-interactions, as credible solitonic candidates for the dark components of disk galaxies at the level of their rotation curves, and they establish \texttt{SpectralPINN} as a practical tool for generating accurate, storage-light solutions amenable to astrophysical forward modeling. With fuller dynamics and a broader data confrontation, this framework can sharpen both astrophysical and particle-physics interpretations of DM at galaxy scales.
%
\acknowledgments

    We would like to thank Ignacy Sawicki and Jorge Castelo Mourelle for their valuable discussions and comments. A. M. Pombo is supported by the Czech Grant Agency (GA\^CR) project PreCOG (Grant No. 24-10780S). This study is supported by the Italian Ministry for Research and University (MUR) under Grant 'Progetto Dipartimenti di Eccellenza 2023-2027' (BiCoQ). 


\bibliographystyle{JHEP}
\bibliography{bibli}

\end{document}